\documentclass[sigconf]{acmart}

\AtBeginDocument{%
  \providecommand\BibTeX{{%
    \normalfont B\kern-0.5em{\scshape i\kern-0.25em b}\kern-0.8em\TeX}}}

\setcopyright{acmlicensed}
\copyrightyear{2023} 
\acmYear{2023} 
\acmDOI{10.1145/3544548.3580701}

\acmConference[CHI '23]{Proceedings of the 2023 CHI Conference on Human Factors in Computing Systems}{April 23--28, 2023}{Hamburg, Germany}
\acmBooktitle{Proceedings of the 2023 CHI Conference on Human Factors in Computing Systems (CHI '23), April 23--28, 2023, Hamburg, Germany}

\acmPrice{15.00}
\acmISBN{978-1-4503-9421-5/23/04}

\usepackage{xcolor}
\usepackage{subcaption}
\usepackage{microtype}
\usepackage{array}
\usepackage{enumitem}
\usepackage{colortbl}
\usepackage{graphicx,calc}
\usepackage{balance}
\usepackage[utf8]{inputenc}
\usepackage{multibib}
\newcites{supp}{Case Example List}

\newcolumntype{\$}{>{\global\let\currentrowstyle\relax}}
\newcolumntype{^}{>{\currentrowstyle}}

\def\markup{0}
\if\markup 1
\usepackage{soul}
\newcommand{\rv}[1]{{\leavevmode\color{blue}#1}}
\else
\newcommand{\rv}[1]{#1}
\newcommand{\st}[1]{}
\fi

\definecolor{symbolc}{HTML}{E60012}
\definecolor{camerac}{HTML}{28A7E1}
\definecolor{soundc}{HTML}{23AC38}
\definecolor{bangc}{HTML}{EA5413}
\definecolor{footagec}{HTML}{F8B62B}
\definecolor{endingc}{HTML}{956134}

\definecolor{fullc}{HTML}{4081BC}
\definecolor{exclamationc}{HTML}{FF681D}
\definecolor{questionc}{HTML}{FFC91D}
\definecolor{ellipsisc}{HTML}{40916A}

\newlength\myheight
\newlength\mydepth
\settototalheight\myheight{Xygp}
\begin{document}

\title{Is It the End? Guidelines for Cinematic Endings in Data Videos}

\author{Xian Xu}
\affiliation{%
\institution{The Hong Kong University of Science and Technology}
\city{Hong Kong}
\country{China}
}
\email{xxubq@connect.ust.hk}

\author{Aoyu Wu}
\affiliation{
\institution{The Hong Kong University of Science and Technology}
\city{Hong Kong}
\country{China}
}
\email{awuac@connect.ust.hk}

\author{Leni Yang}
\affiliation{
\institution{The Hong Kong University of Science and Technology}
\city{Hong Kong}
\country{China}
}
\email{lyangbb@connect.ust.hk}

\author{Zheng Wei}
\affiliation{%
\institution{The Hong Kong University of Science and Technology}
\city{Guangzhou}
\country{China}
}
\email{zwei302@connect.hkust-gz.edu.cn}

\author{Rong Huang}
\affiliation{%
\institution{The Hong Kong University of Science and Technology}
\city{Guangzhou}
\country{China}
}
\email{hr316811369@gmail.com}

\author{David Yip}
\affiliation{%
\institution{The Hong Kong University of Science and Technology}
\city{Guangzhou}
\country{China}
}
\email{daveyip@ust.hk}
\authornote{Corresponding author}

\author{Huamin Qu*}
\affiliation{%
\institution{The Hong Kong University of Science and Technology}
\city{Hong Kong}
\country{China}
}
\email{huamin@cse.ust.hk}
\renewcommand{\shortauthors}{Xu et al.}

\begin{abstract}
Data videos are becoming increasingly popular in society and academia.
Yet little is known about how to create endings that strengthen a lasting impression and persuasion.
To fulfill the gap,
this work aims to develop guidelines for data video endings by drawing inspiration from cinematic arts.
To contextualize cinematic endings in data videos,
111 film endings and 105 data video endings are first analyzed to identify four common styles using the framework of ending punctuation marks.
~\rv{We conducted expert interviews (N=11) and formulated 20 guidelines for creating cinematic endings in data videos.
To validate our guidelines, we conducted a user study where 24 participants were invited to design endings with and without our guidelines, which are evaluated by experts and the general public. The participants praise the clarity and usability of the guidelines, and results show that the endings with guidelines are perceived to be more understandable, impressive, and reflective.}
\end{abstract}


\begin{teaserfigure}
  \includegraphics[width=.9
  \textwidth]{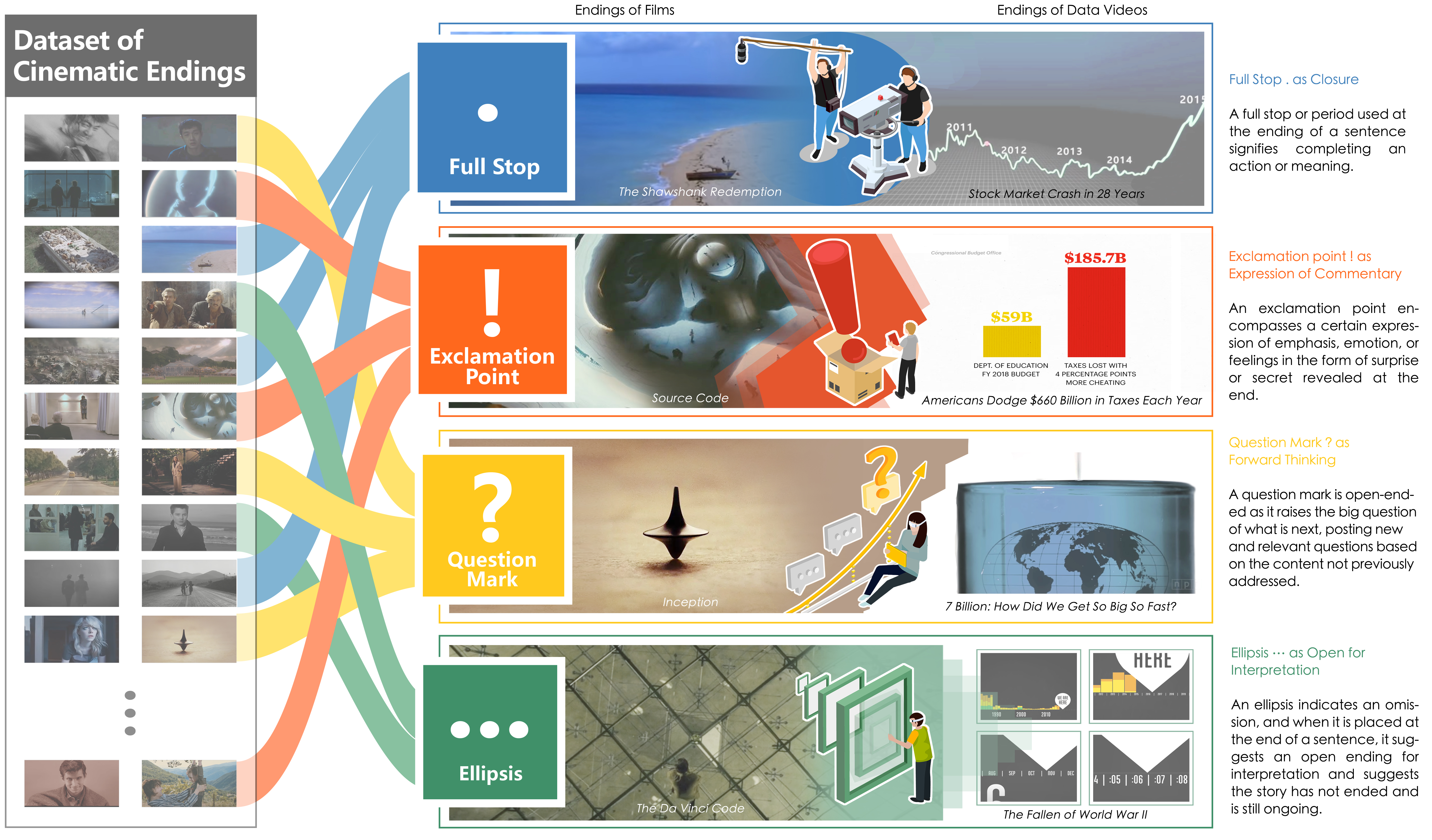}
 \caption{\rv{Four cinematic styles are adaptable to data video endings. The form and content of 111 film endings (left: Dataset of Cinematic Endings) and 105 data video endings are analyzed to identify four styles of data videos endings (right: four cinematic ending definitions) using the framework of punctuation marks, namely, \textit{full stop}, \textit{exclamation point}, \textit{question mark}, and \textit{ellipsis}. Each cinematic ending is shown through a film ending example and a data video ending example (middle).}}
 \Description{This figure shows that four cinematic styles are adaptable to data video endings. The form and content of 111 film endings (left: Dataset of Cinematic Endings) and 105 data video endings are analyzed to identify four styles of data videos endings (right: four cinematic ending definitions) using the framework of punctuation marks, namely, \textit{full stop}, \textit{exclamation point}, \textit{question mark}, and \textit{ellipsis}. Each cinematic ending is shown through a film ending example and a data video ending example (middle).}
  \label{fig:punctuations}
  \vspace{-10px}
\end{teaserfigure}


\maketitle

\section{Introduction}
Data videos, defined as custom motion graphics combining visual and auditory stimuli to promote a data story~\cite{amini2015understanding,segel2010narrative},
are becoming increasingly popular in society and academia.
They are widely shared and viewed on social platforms to communicate data insights and knowledge.
For example,
the data videos by Neil Halloran~\cite{neil:fallen,neil:simulation} have been watched more than 10M times on YouTube.
Leading media outlets such as ~\textit{The New York Times}, ~\textit{The Guardian}, and ~\textit{Vox} also increasingly craft data videos to broadcast information to broad audiences (e.g.,~\cite{Vox,nyt,guardian}).

The increasing prominence of data videos has sparked research interest in investigating how to design compelling, effective data videos.
Much research has focused on analyzing data videos in the wild to distill the design space of visual effects such as animation types~\cite{amini2016authoring} and animated transition designs~\cite{tang2020narrative,shi2021communicating}.
In addition to visual effects, another line of research studies the narrative structures of data videos, such as Cohn's visual narrative structure~\cite{cohn2013visual}~(i.e.,~\textit{Establisher}, ~\textit{Initial}, ~\textit{Peak}, and~\textit{Release})~\cite{amini2015understanding} and Freytag's pyramid structure~\cite{freytag1908freytag}~(i.e.,~\textit{Setting},~\textit{Rising-Climax}, and~\textit{Resolution})~\cite{pyramid2021}.
Nevertheless, how to create effective endings of data videos, vital for understanding, impression, and reflection of a theme at the end of a story, remains unclear.
Often, the author of a data video would appear at the end to address the concluding statement verbally. 
Besides emphasizing key data patterns, the author could also make a commentary or express personal views behind the facts as a researcher, forward thinker, policy advocate, or all of the above~\cite{heyer2020pushing}.
~\rv{However, data visualization, including data videos, is not neutral. Previous work pointed out that designers have affective goals. For example, designers try to call to action at the end of data videos~\cite{lee2022affective}.
Most aesthetic narratives engage emotions, especially for narrative endings~\cite{carroll2007narrative}.}
~\rv{Gerson and Page~\cite{gershon2001storytelling} noted that,~\textit{"The ancient art of storytelling and its adaptation in film and video can now be used to efficiently convey information in our increasingly computerized world."}}
Therefore,
for more lasting impressions and stronger persuasion, thoughts and feelings should be conveyed powerfully with literal, visual, or even cinematic expressions~\cite{yip2020invisible}.
\rv{Studying ending} is challenging as~\rv{it} is always related to the beginning and middle parts of a coherent story in content and context. 
As the author and literature professor James Plath~\cite{James} says, 
~\textit {``The best endings resonate because they echo a word, phrase, or image from earlier in the story, and the reader is prompted to think back to that reference and speculate on a deeper meaning.''}
Form and content are inseparable in an ending, in which all the narrative and visual elements come together to create the conclusion.
Thus, the ending cannot be summarized by its form of visual presentation alone.
Data storytellers need guidelines that integrate content and form in designing data video endings.

To fill this gap,
one approach is to draw inspiration from films, the most mature audio-visual language technology with more than 100 years of history~\cite{segel2010narrative}.
Visual expression has many common forms, such as camera shot and movement, color, sound and music, and editing that both filmmakers and data storytellers use. 
Thus,
some of the styles and techniques filmmakers do to tell their stories could be a source of inspiration for data storytellers to visualize the form, shape, and movement of the data graphics in relation to their space, time, and other data attributes~\cite{yip2020visual}. 


\begin{figure*}
  \includegraphics[width=\textwidth]{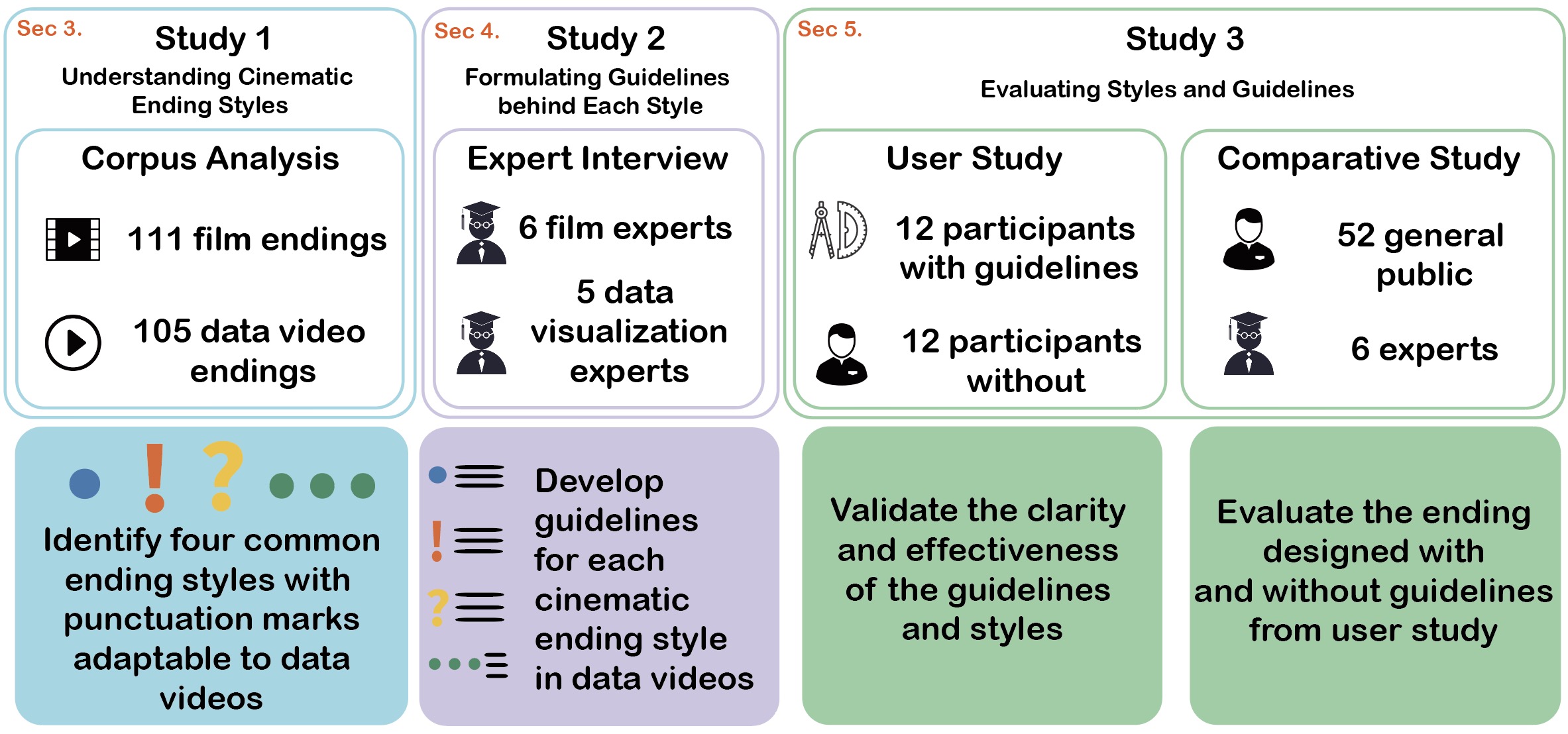}
  \caption{\rv{A survey of corpus analysis, an expert interview, and an evaluation are conducted in this work. The main objective of this research is to identify common cinematic styles and produce a set of guidelines for data designers to create endings with cinematic styles. In Study 1, a large corpus of films and data videos is explored to propose common ending styles adaptable to data videos. Four common styles represented by four punctuation marks, as well as the classic ending examples of films and data videos under each punctuation mark, which are used as materials for the next expert interviews, are classified. In Study 2, the guidelines behind each ending style are derived from expert interviews. In Study 3, a user study and a comparative study are conducted to validate the usability and effectiveness of the guidelines and ending styles.}}
  \label{fig:method}
   \Description{This figure shows an overview of the methodology in our work. A survey of corpus analysis, an expert interview, and an evaluation are conducted in this work. The main objective of this research is to identify common cinematic styles and produce a set of guidelines for data designers to create endings with cinematic styles. In Study 1, a large corpus of films and data videos is explored to propose common ending styles adaptable to data videos. Four common styles represented by four punctuation marks, as well as the classic ending examples of films and data videos under each punctuation mark, which are used as materials for the next expert interviews, are classified. In Study 2, the guidelines behind each ending style are derived from expert interviews. In Study 3, a user study and a comparative study are conducted to validate the usability and effectiveness of the guidelines and ending styles.}
\end{figure*}

In this work,
how to create effective endings of data videos with cinematic styles is studied.
To inform our research,~\rv{this work refers to previous work in data video structures~\cite{amini2015understanding, cao2020examining}, and} the ending of a data video is defined as \textit{the content and form of the final scene or moment of the story at the ending of data videos.}
~\rv{\autoref{fig:method} is an overview of the methodology in our work.}
First, the endings of 111 masterpiece films and 105 data videos are analyzed.
Studying the complexity of boundless forms of content and visual styles finds that using four common punctuation marks to end a sentence can visualize the different tones of an ending, which can alter the final meaning of a sentence and help illustrate the complex and dynamic relationships between content and form. 
In ~\autoref{fig:punctuations},
these punctuation marks provide the ending its last tone of voice, which defines the style and meaning of the ending. 
Then, an interview study is conducted with 11 participants from diverse backgrounds, including cinematographers, directors, film experts, and data visualization researchers.
The interview study results in a set of 20 guidelines.
To evaluate our guidelines,
~\rv{24 participants are further recruited to create data video endings with and without the guidelines, which six experts and 52 general audiences then rate}. 
The results show that participants can efficiently and effectively follow the guidelines to create a more understandable, impressive, and reflective ending. 
The endings with guidelines are also rated as more creative and cinematic by the experts. 
All of the participants praise the clarity and inspiration of the guidelines.
In conclusion,
our contributions are as follows:
\begin{itemize}[leftmargin=*]
    \item Four types of punctuation mark endings with cinematic styles adaptable to data videos are presented by analyzing 111 films and 105 data videos. 
    \item Twenty guidelines are derived by interview studies with six film experts and five data visualization experts.
    \item The effectiveness and usefulness of the guidelines for creating cinematic ending are demonstrated through a user study and a comparative study.
\end{itemize}

\section{Related Work}
Our work is related to narrative visualization, data videos, and \rv{narrative} endings.

\subsection{Narrative Visualization}
The concept of narrative visualizations regarding the integration of data graphics into storytelling was formally introduced by Segel and Heer~\cite{segel2010narrative} in 2010. 
They identified seven common genres for narrative visualizations,
including \textit{magazine style, annotated
chart, partitioned poster, flow chart, comic strip, slide show, and film/video/animation.}
Since then,
researchers have explored additional genres such as \textit{data comics}~\cite{bach2018design, zhao2019understanding, kim2019datatoon, wang2021interactive}, \textit{timeline visualization}~\cite{brehmer2016timelines}, and \textit{data GIFs}~\cite{shu2020makes}.
Moreover, many studies deeply explored and studied the connection between some human perception concepts and visualization, such as visualization recognition and recall~\cite{borkin2015beyond}, embellishments in visualization~\cite{borgo2012empirical}, and anthropographics and visualization~\cite{boy2017showing, morais2020showing}.
Recently, Coelho et al.~\cite{coelho2020infomages} ~\rv{proposed ``infomages'' --- that embedded common data charts into thematic images that are related to the subjects of data (e.g., embedding a pie chart into a laptop in an image of a hacker using the laptop to represent the number of hacking incidents by region).}

In addition, 
researchers seek to provide suggestions and guidelines on effective data storytelling from an interdisciplinary perspective~\cite{gershon2001storytelling}.
Many studies conduct experiments to surface psychological effects in data visualizations, such as the framing effect~\cite{hullman2011visualization}
and the curse of knowledge~\cite{xiong2019curse}.
Another line of research studies narrative visualization from the perspective of literature~\cite{lee2015more, amini2015understanding, pyramid2021}.
For example,
\rv{Lee et al.~\cite{lee2015more} derived the data storytelling process from data journalism literature and pointed out that the visualization community still lacks enough attention to the structure and sequence of compelling story pieces, including a beginning and an ending, which influences the reception of data stories.}
\rv{Beginnings and endings are the necessary terms in a narrative structure~\cite{cao2020examining}. However, existing research had more exploration about the beginnings of narrative visualization~\cite{segel2010narrative, xu2022fromwow, pyramid2021, amini2015understanding}. For instance, Segel and Heer~\cite{segel2010narrative} explored and discussed the design patterns of narrative visualization, which paid more attention to the beginning parts of the interaction of the story.
Nonetheless, the connection between beginnings and endings was noted in these works. For example, Yang et al.~\cite{pyramid2021} identified one of the narrative patterns for data story ending as \textit{Echoing the beginning}. Xu et al.~\cite{xu2022fromwow} derived one of the opening styles as \textit{Ending First}, which presented associating the opening and the ending as a common nonlinear narrative technique. As such, creating the data story ending can inspire its opening.
To the best of our knowledge, the endings in the community of HCI and VIS that are important and worthy of further research remain much underexplored. Little related work can be found: Amini et al.~\cite{amini2015understanding} decomposed the narrative structure of data videos into four categories and identified the importance of the ending as \textit{Release} because it takes more time to design, and concluded that the \textit{Release} contains~\textit{New Fact, Repetition, Take Away, and Question}. Yang et al.~\cite{pyramid2021} further decomposed the \textit{Release} into four~\rv{narrative patterns (namely, \textit{Recap}, \textit{Predicting the future}, \textit{Echoing the beginning}, and \textit{Next steps}).}}
Compared with their research,
this work aims to study the common styles of data story endings that have been underexplored.
Specially,
a cinematic lens is taken by summarizing styles in film endings and analyzing how they might be applied to data videos.

\subsection{Data Videos}
As a popular genre for storytelling,
data videos have gained intense research interest over the last decade.
Researchers have developed a wide range of tools to ease the creation.
For example,
DataClips~\cite{amini2016authoring} allowed nonexperts to create and assemble data-driven clips.
Gemini~\cite{kim2020gemini} and Canis~\cite{ge2020canis} provided declarative grammar to create animations.
\rv{Recently, Shin et al.~\cite{shin2022roslingifier} introduced a new genre of data-driven storytelling, namely, \textit{data presentation}, similar to data videos, and presented an approach of automatically generating a data fact, enabling users to improve the stories and produce animated visualization. Sun et al.~\cite{sun2022erato} demonstrated a method for producing data facts and visuals using the interpolation algorithm, which helps create data stories more efficiently through human-machine cooperation. Although these existing tools help (semi)-automatically create data stories, it still requires more studies to provide diverse visual and content designs to enrich the design templates and appropriate recommendations from authoring tools.}
In addition to authoring tools,
researchers have sought to distill the design space and propose design guidelines.
Much research focuses on low-level visual effects such as animation types~\cite{amini2016authoring} and animated transition designs~\cite{tang2020narrative,shi2021communicating}.
Recent work further links visual effects with audience feedback, such as affection and attractiveness.
For example,
Lan et al.~\cite{lan2021kineticharts, lan2022negative} studied how animation could evoke different emotions in data videos.
Xu et al.~\cite{xu2022fromwow} summarized six common styles of video openings and formulated 28 guidelines to create an attractive opening with cinematic effects. 
The above study is extended by examining the endings of data videos with cinematic styles.
Critically,
the ending is content-related and thus cannot be studied by examining its visual representation alone.
This characteristic makes this study challenging and different from previous ones that largely focus on visual presentations but pay slight attention to narrative content.



\rv{\subsection{Narrative Endings}
Many researchers have studied narrative endings in literature and film. 
Early in the development of narratives, researchers have \nobreak noticed the important role that endings can play. Endings are considered a remarkable feature of narratives~\cite{smith1968poetic, carroll2007narrative, abbott2021cambridge}. 
To better understand the role of endings in the narrative, prior research has defined and discussed endings in different fields (e.g., poem~\cite{smith1968poetic}, novel~\cite{adamo1995beginnings}, and film~\cite{neupert1995end}) for a long study history. In terms of literature, Aristotle first proposed the ending concept in narrative and aimed to anatomize and conclude the precisely appropriate place for endings representation~\cite{aristotle2006poetics}.
Klauk et al.~\cite{klauk2016empirical} divided narrative endings into seven categories to understand the features of the text in narrative endings, which helped explore reader reaction to the narrative endings.
Noel Carroll further discussed that the ending is the result of all the pressing questions having been answered and a key element to make the story exciting. Specifically, to ensure impression of complete narrative, endings provide necessary conditions for some scenes of the stories such as the establishing scenes of a film or introductory paragraphs for a\break novel~\cite{carroll2007narrative}.}

\rv{Similar to the popular discussion on the endings of literature, }
film endings are a cinematic art form that~\rv{attracts much} attention in the film industry and filmmaking research~\cite{giannetti1999understanding, kunkle2016cinematic}.
Related research starts with concluding types of endings in films such as open and closed, happy, and sad~\cite{neupert1995end, preis1990not}. 
The unlimited potential in endings fueled films across the ages.
For example, films such as~\textit{Citizen Kane (1941)}~\cite{citizen}, \textit{Bicycle Thief (1948)}~\cite{bicycle}, and \textit{The 400 Blows (1959)}~\cite{400blows} are considered an artistic cornerstone in the history of the cinema for their open endings~\cite{cutting2016narrative}.
Previous work found that film endings could create and alter the meanings of the stories and the experiences of viewers~\cite{stone2002hope}.
Research in film has also been exploring how to create better film endings.
For example, King~\cite{king2007don} compared the effects of traditional endings and teaser endings on viewers' enjoyment of horror films and found that participants prefer traditional ones.
Berys et al.~\cite{gaut2010philosophy} discussed how narrative techniques in films could affect the audience's interpretations of an ending.
Research that systematically summarizes cinematography or narrative techniques at the ending of films is lacking, but those of discussions in the wild are many. 
Thus, we identified film techniques by analyzing 111 films and expert interviews, and we selected ones adaptable to data videos in this work.
\rv{After comparing film endings with data video endings, we found that they both help to convey insights from two perspectives. First, closed endings could help summarize and recap the essential insights. Second, open endings prompt the audience to think about the next step of the story or interact with the next data visualization, which could be useful in the ending designs of interactive visualization.
Thus, this paper investigates how to adapt narrative ending techniques, especially cinematic ending techniques, to data videos. To the best of our knowledge, our work is the first one that systematically explores the combination of cinematic storytelling and data video\break endings.}

\rv{\section{Study 1: Understanding Cinematic Endings Styles}}
\rv{To contextualize cinematic endings to data videos, a survey is conducted to explore a large corpus of films and data videos. We identified four types of punctuation marks with cinematic ending styles applicable to data videos. In this section, we first describe our methodology for identifying cinematic endings styles, and then elaborate on these ending styles through cinematic ending examples from films and data videos.}




 

\subsection{Methodology}
An iterative coding approach is adopted to analyze films and data videos.

\textbf{Dataset.}
Our dataset consists of 111 films and 105 data videos.
For films,
this work refers to the YouTube video \textit{Best Movie Endings of All Time} by CineFix - IGN Movies and TV~\cite{CineFix}.
CineFix is home to informational and entertaining movie and TV show content made by experts for the modern-day cinephile, and has about 3.8 million subscribers. 
It produces a comprehensive review of over 100 film endings for our reference and has generated more than 318k views, 9k likes, and 1.4k comments since May 1st, 2021. 
Some films with cinematic endings were professional inputs from our research team member, who has about 25 years of experience in film teaching and producing.
For data videos, high-quality data videos were collected from previous narrative visualization studies~\cite{shi2021communicating,pyramid2021,xu2022fromwow,segel2010narrative,amini2016authoring, amini2015understanding}, some popular video platforms, and well-known news agencies that have produced data videos (e.g., YouTube~\cite{youtube}, Tencent Video~\cite{tecent}, Vox~\cite{Vox}, New York Times~\cite{nyt}, The Economist~\cite{economist}).~\rv{In particular, we used keywords such as ``data storytelling,'' ``data stories,'' and ``data-driven videos'' to search for data videos and subsequently process videos with a high number of views. All analyzed cases can be found in the supplemental materials.}

\textbf{Procedure.}
The authors were divided into two groups to analyze those films and data videos.
The film group consisted of a professor with about 25 years of experience in teaching films and another author with eight years of experience in film writing and producing film.
The data video group included three authors with research experience in data visualization and human computer interaction.
~\rv{Two groups adopted thematic analysis~\cite{braun2006using} to code the ending styles of data videos and films separately, and held meetings twice a week during the coding to bridge the gap between data videos and films.
The goal is to find a set of styles for classifying and describing the endings of films and data videos.
To theorize the classification,
the groups discussed and summarized the characteristics of film and data video endings in terms of content and form for each style.}
\rv{Initially, our classification of ending styles was inspired by the literature on film endings that use punctuation marks to describe cinematic endings in films~\cite{schulz, brody2008punctuation, kunkle2016cinematic, stone2002hope, editing}.
First, three common ending punctuation marks~\cite{14punctuation, guide} were used, namely, \textit{full stop}, \textit{exclamation point}, and  \textit{question mark}.
During iterative coding, another common punctuation mark, the \textit{ellipsis}, can also be used at the end of a sentence to imply omission.
Although \textit{ellipses} need to be formally written as …. or …? or …!~\cite{nunberg1990linguistics},
they are considered a common film style that implicitly expresses a theme or an idea~\cite{ryan2015heretical, filmsite, filmnarrative, hollywoodlexicon, art}.
Thus, this punctuation mark was added to our analysis framework.
After rounds of coding and discussion, new ending styles were no longer found toward the end of analyses. This result suggested that our codes almost reached saturation, and the groups stopped analyzing new videos. Differences were resolved by discussions. Finally, the groups reached a consensus on four cinematic endings styles using four punctuation\break marks.}





\rv{\subsection{Results and Analysis}}
In the following text,~\rv{our results on four cinematic ending styles are reported and elaborated using the framework of punctuation marks, namely, \textit{Full Stop as Closure}, \textit{Exclamation Point as Expression of Commentary}, \textit{Question Mark as Forward Thinking}, and \textit{Ellipsis as Open for Interpretation}.}
For each style, our analysis and organization are based on a structured format, including the definition, the representative examples in films, and a case study examining data videos through this cinematic lens.

\begin{figure}[!ht]
  \centering
      \includegraphics[width=\linewidth]{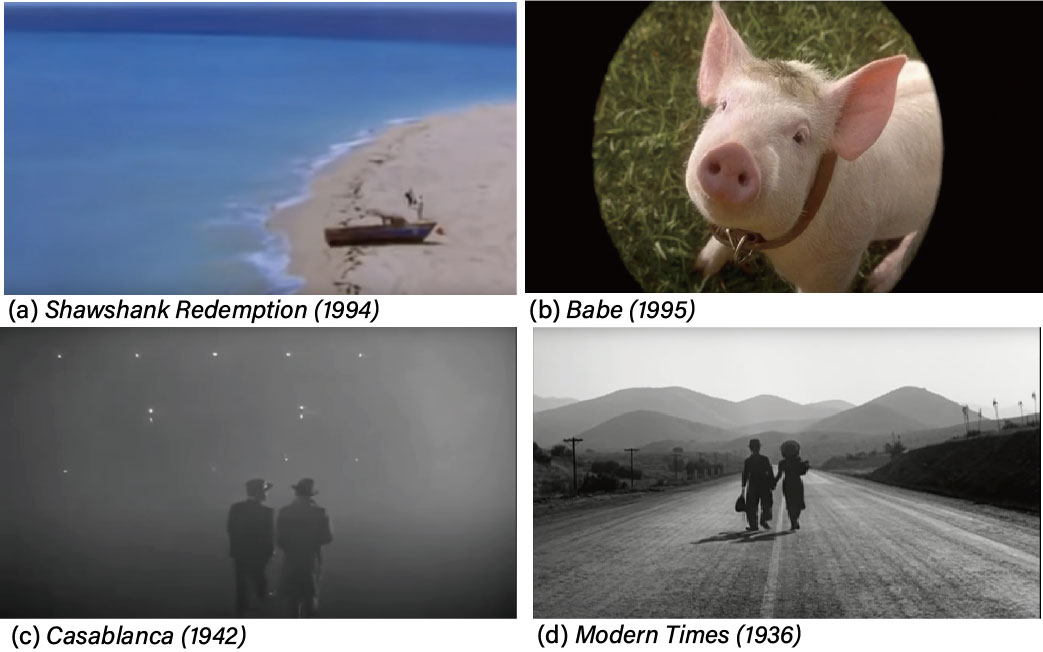}
    \caption[]{\rv{Endings of \textit{The Shawshank Redemption (1994)}~~\protect\cite{shawshank}, \textit{Babe (1995)}~~\protect\cite{babe}, \textit{Casablanca (1942)}~~\protect\cite{casablanca}, and \textit{Modern Times (1936)}~~\protect\cite{modern} using \textit{full stop} with the film techniques of a zoom out camera movement, wide shot, and the iris effect to achieve closure}}
    \Description{This figure shows the endings of four film endings of \textit{The Shawshank Redemption (1994)}~\cite{shawshank}, \textit{Babe (1995)}~\cite{babe}, \textit{Casablanca (1942)}~\cite{casablanca}, and \textit{Modern Times (1936)}~\cite{modern} using full stop with the film techniques of a zoom out camera movement, wide shot, and the iris effect to achieve closure.}
    \label{fig:fullstop}
  \end{figure}

\subsubsection{\textcolor{fullc}{Full Stop as Closure}}
A full stop or period used at the ending of a sentence signifies completing an action or meaning.
Therefore, it functions as a close-ended punctuation mark. 
In the literary sense, this punctuation mark can be interpreted as things that have been resolved or questions that have been answered with closure.
It can also suggest the closure of a causality loop or complete connection between past and present. 
As expected, full stops appeared to be most common in the corpus of films (45/105) and data videos (37/111).

\begin{figure}[!ht]
  \centering
      \includegraphics[width=\linewidth]{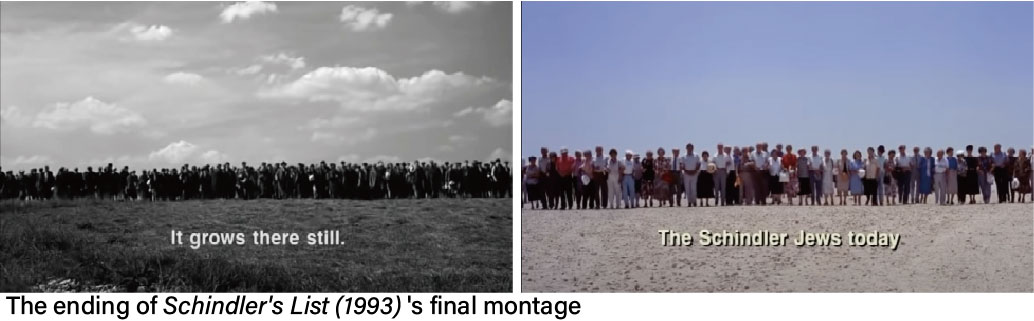}
    \caption{\rv{Ending of \textit{Schindler's List (1993)}~\protect\cite{schindler} using \textit{full stop} through the film technique of montage as the conclusion of the story}}
    \Description{\rv{This figure shows the ending of \textit{Schindler's List (1993)}~\cite{schindler} using \textit{full stop} by the film technique of montage as the conclusion of the story.}}
    \label{fig:fullstop2}
 \end{figure}

\textbf{Films.}
Ending with a zoom out camera movement, wide shot, or montage are examples of cinematic styles that can signify and express the sense of closure and full stop of a film story.
For example, ~\textit{The Shawshank Redemption (1994)}~\cite{shawshank}, as shown in~\autoref{fig:fullstop} (a), ends in a vast zoom out to show a wide shot of the two main characters as free men walking by the sea, a new world that is much larger than their prison cell after years of imprisonment. 
Their freedom is symbolized by this great zoom out and wide shot as the closing shot of the film. 
In some cases, the camera's movements are accompanied by the characters' movements.
In the last shot of~\textit{Casablanca (1942)}~\cite{casablanca}, as shown in~\autoref{fig:fullstop} (c), as the two characters are left to walk away from the camera, the shot size changes from medium to wide as the characters are walking away to begin their ``beautiful friendship." 
Likewise, the ending shot in ~\textit{Modern Times (1936)}~\cite{modern}, as shown in~\autoref{fig:fullstop} (d), shows Charles Chapman is walking away with his girl in a static shot that changes from a medium shot to a wide shot, which signifies the sense of completion or closure in a film story.
Another common editing technique in a full stop ending is the iris effect that serves like a zoom in to close up to focus on the subject, such as the ending shot of~\textit{Babe (1995)}~\cite{babe}, as shown in~\autoref{fig:fullstop} (b), that features the small pig as the final hero of the story. 
Finally, montage is often used to close a film to show the causal relation between action and reaction as the conclusion of the plot. 
For example, the ending of the final montage of~\textit{Schindler's List (1993)}~\cite{schindler} ~(\autoref{fig:fullstop2}) shows the transitions from fiction to reality by showing the real people whom Schindler saved in real life.


\begin{figure}[!ht]
  \centering
  \begin{minipage}[b]{\linewidth}
      \includegraphics[width=\textwidth]{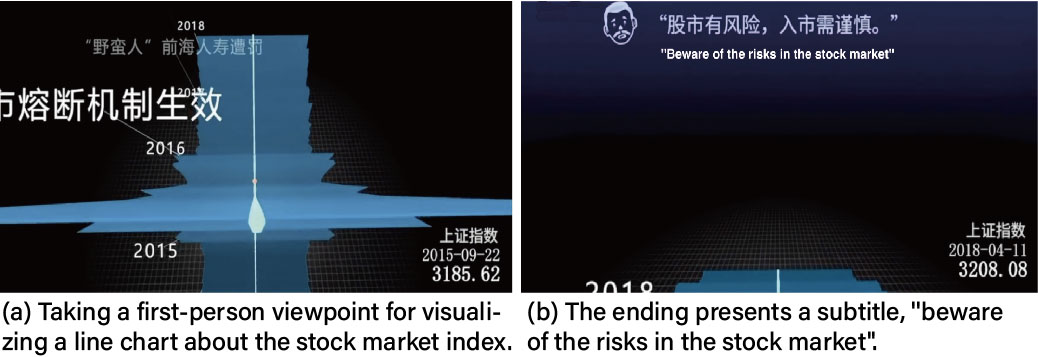}
    \caption{\rv{Data video ending of \textit{Stock Market Crash in 28 years}~~\protect\cite{stock} using \textit{full stop} through providing clear, direct vital messages}}
    \Description{This figure shows the data video ending of \textit{Stock Market Crash in 28 years}~\cite{stock} using \textit{full stop}through providing clear, direct vital messages.}
    \label{fig:fullstop-datavideos}
  \end{minipage}
     \vspace{-20px}
  \end{figure}

\textbf{Data Videos.} The endings of full stop often provide clear, direct vital messages such as the most substantial data pattern, an ultimate suggestion, or a call to action.
However,
they lack a solid, concise summary for emphasizing the main point.
For example,
the data video \textit{Stock Market Crash in 28 years}~\cite{stock}, as shown in~\autoref{fig:fullstop-datavideos} (a), takes a first-person viewpoint for visualizing a line chart about the stock market index,
engaging audiences in experiencing the ups and downs of the stock market on a roller coaster ride.
The video proceeds by switching between different viewpoints to make audiences visually immersed.
Similarly,
it could end with cinematic techniques to render a cogent summary of data and provoke reflections.
For instance, it could apply a smooth transition from the close-up shot at the last time point in the chart to a medium shot to a wide shot to review the overall fluctuating trend of the stock market index.
Instead,
the video only presents a plain subtitle, ``Beware of the risks in the stock market,'' as shown in~\autoref{fig:fullstop-datavideos} (b), so that audiences might overlook the critical data insights.
This case demonstrates that data videos have enormous potential for improvements by learning from film arts,
underscoring the importance of this research.


\subsubsection{\textcolor{exclamationc}{Exclamation Point as Expression of Commentary}}
Full stop and exclamation points are considered closed ended, which signifies the end of a complete action in a sentence. 
Specifically, an exclamation point encompasses a certain expression of emphasis, emotions, or feelings in the form of surprise or secret revealed at the end. 
This type of ending can be impressionistic and let the audience have the last ``wow'' moment or a new level of emotional reaction.
This work finds that exclamation points appeared to be moderately common in the film corpus (29/111), where the suspense was lifted in the film endings.
However, exclamation points were not common in data video endings (11/105) that might inspire future practices.

\begin{figure}[!ht]
  \centering
      \includegraphics[width=\linewidth]{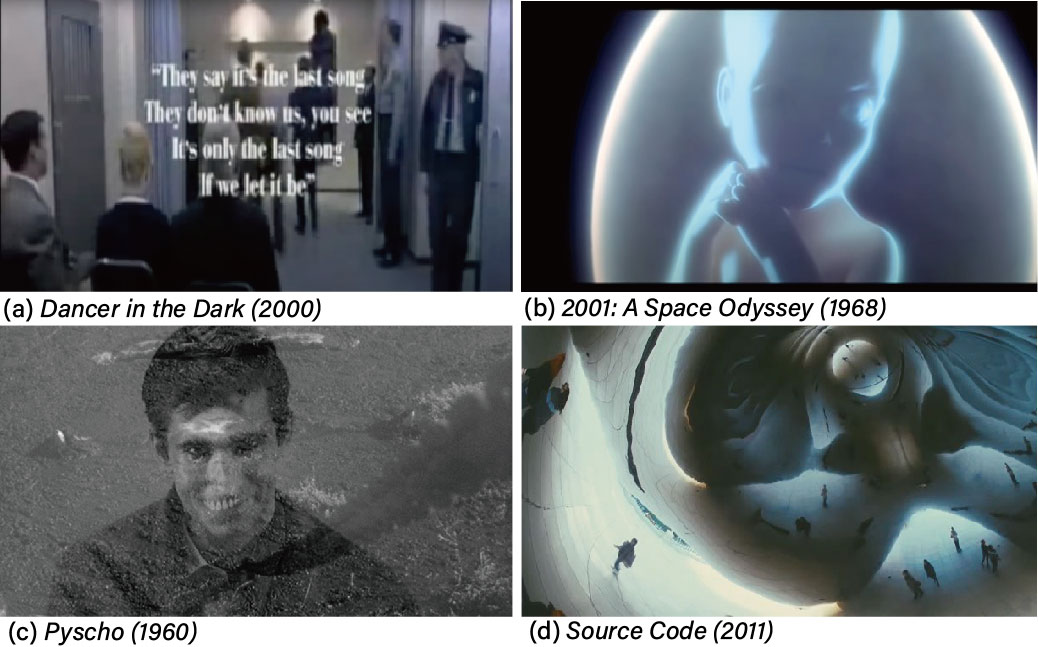}
    \caption{\rv{Endings of \textit{Dancer in the Dark (2000)}~~\protect\cite{dancer}, \textit{2001: A Space Odyssey (1968)}~~\protect\cite{2001}, \textit{Pyscho (1960)}~~\protect\cite{pyscho}, and \textit{Source Code (2011)}~~\protect\cite{soure} using \textit{exclamation points} with stunning montage sequences and multiple collages of different visual shapes to sum up the theme visually}}
    \Description{This figure shows the endings of four films using exclamation points. There are endings of \textit{Dancer in the Dark (2000)}~\cite{dancer}, \textit{2001: A Space Odyssey (1968)}~\cite{2001}, \textit{Pyscho (1960)}~\cite{pyscho}, and \textit{Source Code (2011)}~\cite{soure} with stunning montage sequences and multiple collages of different visual shapes to sum up the theme visually.}
    \label{fig:exclamation}
\end{figure}

\textbf{Films.}
The ending of the exclamation point is a strong cinematic expression of the author's thoughts, emotions, and feelings that appeal to the audience's emotional senses in various forms. 
Sometimes it can appear as a sharp irony in life and death such as the tragic endings in ~\textit{Life is Beautiful (1997)}~\cite{life} and ~\textit{Dancer in the Dark (2000)}~\cite{dancer}, as shown in~\autoref{fig:exclamation} (a). 
~\rv{Sometimes this expression of commentary appeals to our sensory and emotional senses and leaves the audience with a ``wow'' reaction with different interpretations.} 
~\textit{Persona (1966)}~\cite{persona} and ~\textit{2001: A Space Odyssey (1968)}~\cite{2001}, as shown in~\autoref{fig:exclamation} (b) films open and end on visually stunning montage sequences as the visual expression of commentary with their powerful exclamation points at the end. 
In addition to using a montage, the filmmaker can also explain and sum up the theme visually in one shot in the ending scene. 
For example, the plot secret revealed at the end of~\textit{Pyscho (1960)}~\cite{pyscho}, as shown in~\autoref{fig:exclamation} (c) shocked many audiences. 
Combined with the effects of voice-over, sound, and music, one of the haunting last shots of the ending superimposed three layers of images with the smiling killer who has a split personality, the corpse of the mother, and the sunk car with the victim being pulled up by the police. 
Another expressive shot is in the ending of~\textit{Source Code (2011)}~\cite{soure}, as shown in~\autoref{fig:exclamation} (d) in which the filmmaker showed a shot with reflections of different shapes to conclude its theme about multiverses or parallel universes visually.

\textbf{Data Videos.}
Data videos in this category end with data insights that are surprising or unexpected to the audiences. 
A surprise is often made by compassion or contrast,
impressing audiences with a strong affection to strengthen their understanding of data.
Such style is exemplified by~\textit{America Dodge \$660 Billion in Taxes Each Year}~\cite{america}.
Its ending contrasts education funding with taxes, that is, the money lost with four percentage points more cheating is triple the Department of Education's budget for 2018.
This contrast uses the wide public concern on education funding to highlight the severe tax evasion, leaving a strong expression of commentary.
Another kind of contrast happens when a story reveals discordance or incongruity of facts, which uses irony to increase the dramatic effect.
In the video \textit{Weed is not More Dangerous than Alcohol}~\cite{weed}, the narrator questions the Federal government's unfairly stricter restriction on marijuana compared with alcohol despite the evidence that the overdose of alcohol caused the death of over 10000 times that caused by marijuana.  
The ironic mismatch between the level of danger and strength of restrictions challenges the audiences' expectations and motivates them to reflect on current policies. 
However, most of the content was directly delivered by the author talking about them in front of the camera without any visual treatment, indicating the importance of guidelines for cinematic endings.
 
\subsubsection{\textcolor{questionc}{Question Mark as Forward Thinking}}
A question mark is open ended as it raises the great question of what is next, posting new, relevant questions based on the content not previously addressed. It is also associated with unsolved answers or mysteries, often suitable narrative materials that arouse curiosity.
It often opens up new questions or mysteries related to the plot in films.
Question marks were the least common in films (10/111) and data videos (6/105).
Data videos often provide answers to the questions, and new questions are raised at the end.



\begin{figure}[!ht]
  \centering
  \begin{minipage}[b]{\linewidth}
      \includegraphics[width=\linewidth]{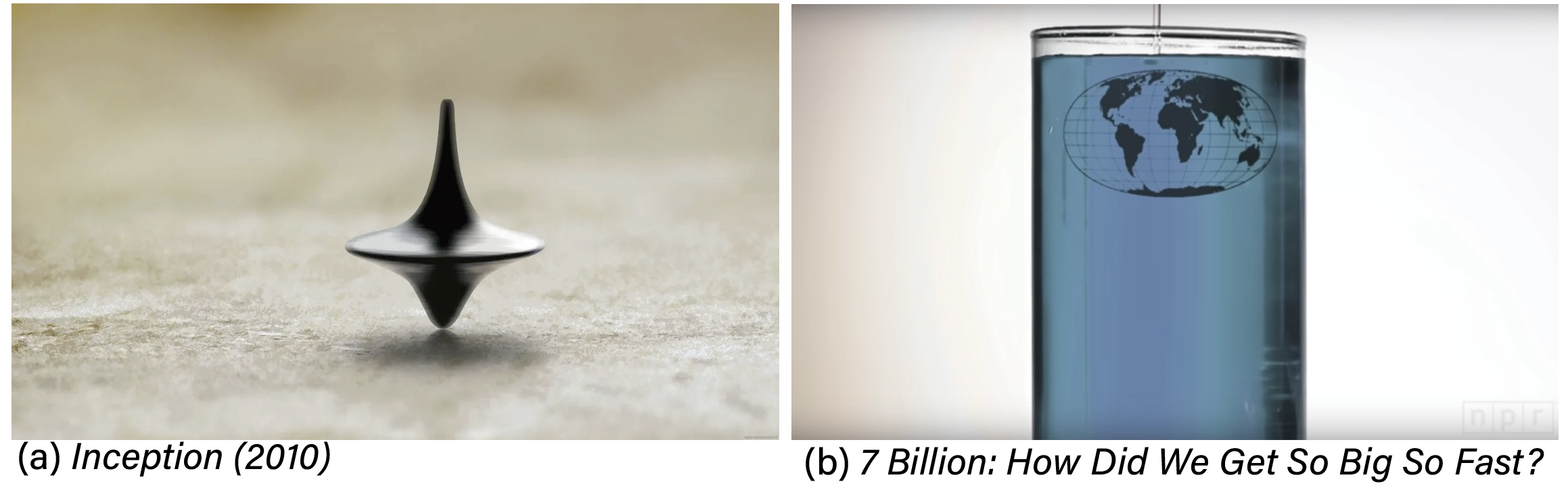}
    \caption{\rv{Endings of \textit{Inception (2010)}~~\protect\cite{inception} and \textit{7 Billion: How Did We Get So Big So Fast?}~~\protect\cite{7billion} using \textit{question marks} with a sudden change or a break in pattern}}
    \Description{This figure shows the endings of \textit{Inception (2010)}~\cite{inception} and \textit{7 Billion: How Did We Get So Big So Fast?}~\cite{7billion} using \textit{question marks} with a sudden change or a break in pattern.}
    \label{fig:question}
  \end{minipage}
  \end{figure}

\textbf{Films.}
In time-based media, timing or tempo is everything. Selecting a precise moment to end a shot can create new meaning or raise the relevant question to the audience.
The ending shot of the famous film \textit{Inception (2010)}~\cite{inception}, as shown in~\autoref{fig:question} (a), has raised discussions for many years. 
The unusually long shot on the spinning top kept spinning endlessly as if it suggested the main character was still in his dream state, but after many seconds before the spinning top appeared to be wobbling, the film ended. 
This ending style of showing the unusual property of a symbolic object has made a strong impression on the audience. The filmmaker has invited his audience to think about what is real or not with this ending shot. 
Another favorite question to ask in drama is what is next in the story. A sudden change or a break in pattern always draws attention. 
In the final scene of the famous classic film \textit{The Graduate (1967)}~\cite{graduate}, when the main character runs away with the bride, the two young lovers laugh, but as they jumped on the city bus, their facial expressions turned blank. This sudden change of tone and pace through the characters' facial expressions raises a question about their future. 
  
\textbf{Data Videos.}
Indeed, the author would appear at the end of a data video to raise new questions directly, which can be found in many data videos.
However, these questions are often expressed verbally and directly without using visual styles.
Perhaps adding a visual scene at the end could leave a stronger impression on the audience.
Data graphics could function like characters in a story.
Change or break in the form, such as color, shape, or movement, can signify a change in meaning at the end. 
The data video \textit{7 Billion: How Did We Get So Big So Fast?}~\cite{7billion} is an exception that has a cinematic ending similar to \textit{Inception (2010)}~\cite{inception}, as shown in~\autoref{fig:question} (a).
In~\autoref{fig:question} (b),
its ending shots symbolize the world's population growth by filling the glass with water, that is, when the water glass becomes full,
our earth reaches the carrying capacity.
The last frame uses a similar ending style of ending abruptly.
This visual metaphor immediately raises the question: Will the water spill over? Could the earth accommodate the rapid population growth? 
This case illustrates how animation and symbolism offer a combined power to evoke questions cinematically.



\subsubsection{\textcolor{ellipsisc}{Ellipsis as Open for Interpretation}}
An ellipsis indicates an omission, and when placed at the end of a sentence, suggests an open ending for interpretation and suggests the story has not ended and is still ongoing.  
This category also applies to an ending that is up to the audience to make their meaning out of the intended ambiguity.
Ellipses are the second most common style in the film (35/111) and data video (43/105) corpus.
It emphasizes that the story keeps evolving or the data are still changing,
thus provoking audiences' thoughts.



\begin{figure}[!ht]
  \centering
  \begin{minipage}[b]{\linewidth}
      \includegraphics[width=\linewidth]{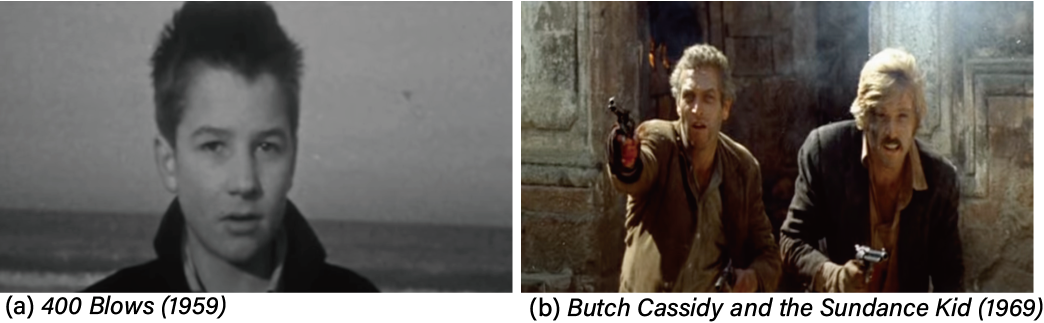}
    \caption{\rv{Endings of \textit{400 Blows (1959)}~~\protect\cite{400blows} and \textit{Butch Cassidy and the Sundance Kid (1969)}~~\protect\cite{butch} using \textit{ellipses} with the freeze-frame shot}}
    \Description{This figure shows the endings of \textit{400 Blows (1959)}~\cite{400blows} and \textit{Butch Cassidy and the Sundance Kid (1969)}~\cite{butch} using \textit{ellipses} with the freeze-frame shot.}
    \label{fig:ellipsis}
  \end{minipage}
\end{figure}

\begin{figure}[!t]
  \centering
  \begin{minipage}[b]{\linewidth}
      \includegraphics[width=\linewidth]{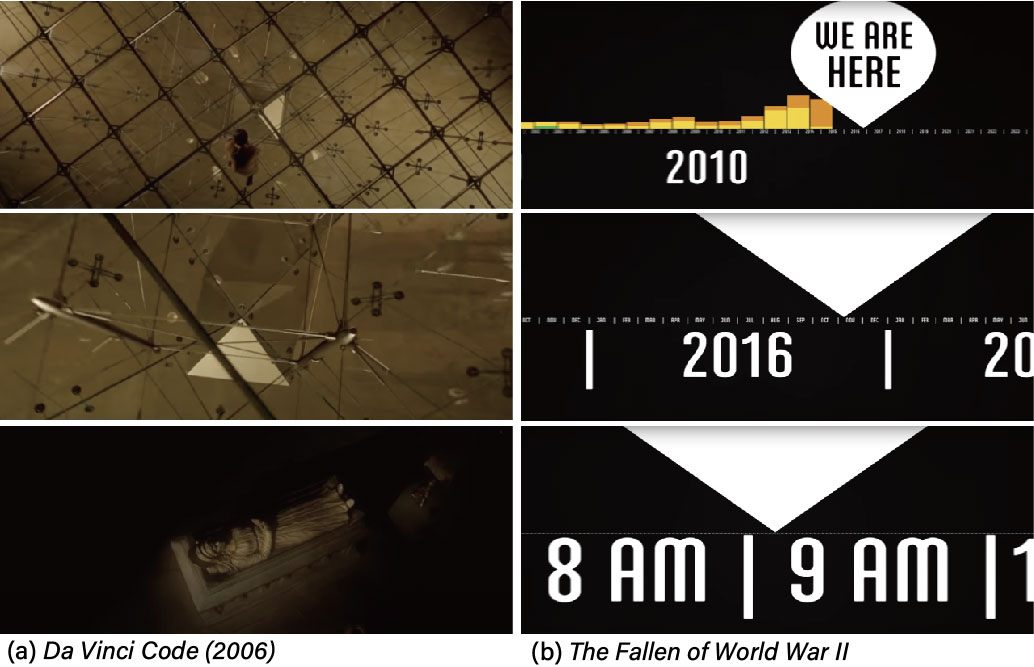}
    \caption{The endings of \textit{The Da Vinci Code (2006)}~~\protect\cite{davinci} and \textit{The Fallen of World War II}~~\protect\cite{neil:fallen} using \textit{ellipses} through a long zoom in shot}
    \Description{This figure shows the endings of \textit{The Da Vinci Code (2006)}~\cite{davinci} and \textit{The Fallen of World War II}~\cite{neil:fallen} using \textit{ellipses} through a long zoom in shot.}
    \label{fig:ellipsis-datavideos}
    \vspace{-10px}
  \end{minipage}
  \end{figure}

\textbf{Films.}
When an ending is ambiguous and offers no clear solution or resolution, it fits the description of the ellipsis. This open-ended style can mean that the story might end here, but life continues, or the problem has not been solved. The famous freeze-frame of the last shot of \textit{400 Blows (1959)}~\cite{400blows}, as shown in~\autoref{fig:ellipsis} (a) and \textit{Butch Cassidy and the Sundance Kid (1969)}~\cite{butch}, as shown in~\autoref{fig:ellipsis} (b), have been the subject of many film analyses.
This style gives the audience opportunities to draw their conclusions and have different interpretations of the story. 
Besides a freeze-frame, the last shot could end before the result or action is completed or delivered.
Similarly, in the ending scene of \textit{The Da Vinci Code (2006)}~\cite{davinci}), as shown in~\autoref{fig:ellipsis-datavideos} (a), a long zoom in was used to show that the protagonist found out the Magdalene stone pavilion was hidden under the pyramids of conjecture. 
The film is over, but the conjecture still needs to continue.
Similarly, the ending shot of \textit{Ex Machina (2014)}~\cite{machina} is when the AI is disappearing into the crowd with the major crisis still in progress and unresolved. 

\textbf{Data Videos.}
Neil Halloran’s data videos deal with subject matters from war to climate change, but all their endings project the filmmaker’s feelings and concerns about the uncertain future open for interpretation. 
His voiceover expresses that there are lessons to be learned from the data as human development is always continuous. 
This theme is also expressed visually through his cinematic data visualization. One example is the long zoom in and zoom out in the endings of \textit{The Fallen of World War II (2016)}~\cite{neil:fallen}, as shown in~\autoref{fig:ellipsis-datavideos} (b) and \textit{Simulation of a Nuclear Blast in a Major City (2020)}~\cite{neil:simulation}, respectively. 
In the former video, the director not only expresses his view through his voice-over but also visualizes through magnifying the graphical data in a long zoom in shot to suggest that the meanings of peace can be found ``between the lines.''
Similarly, in the latter example, the long zoom out shows the possible destruction of a nuclear blast as the voiceover states, ``We can continue ensuring that this nightmare simulation never becomes a reality.'' 
Both endings let us interpret for ourselves the uncertainty that lies ahead, and they are graphically visualized through cinematic language and expression.
The filmmaker also expresses his personal feelings behind the data at the end. 
His feelings are not defined and expressed in the form of forceful exclamation or question marks, as his forward-thinking view of uncertainty is softly expressed as an ellipsis statement. 
His theme about human development as a continuity of uncertainty is consistently expressed in his endings. 
Therefore, these endings function as more ellipsis marks than exclamation or question marks that usually carry a stronger tone of voice and certainty. 

\subsection{Discussion}
In short, based on previous works~\cite{pyramid2021, amini2015understanding} related to narrative ending styles, our study further emphasizes the contextual relation between the ending and content of the data story. Our work echoes previous narrative elements (e.g., questions, recap, echoing the beginning) and further proposes new cinematic elements (e.g., montage, wide shot, and the iris effect) as discussed in case examples. 

Aside from inspirations drawn from cinematic language for the effective ending, one important feature of our study is to emphasize the design of an ending is constrained and contextualized by its content and how the desired tone can be expressed with a specific cinematic style that aims to create a lasting impression. The chosen ending style does not exist in isolation or independent from the context and content of the work. This study does not just name these elements but builds on and extends them for the cinematic endings. We further incorporate these elements with design objectives and specific guidelines to create a more lasting impression ending through our guidelines.

\rv{\section{Study 2: Formulating Guidelines}}

\rv{Based on the corpus analysis, expert interviews were conducted to derive guidelines for producing cinematic endings for data videos. Specifically, the classic ending examples of films and data videos under four common styles are used as materials for the expert interviews. 
In this section,
our methodology and the resulting guidelines are discussed.}

\subsection{Methodology}
Inspired by~\cite{kim2022mobile, xu2022fromwow, lan2022negative}, experts from different backgrounds were interviewed to derive the guidelines.
First, film experts were asked about suggestions for applying cinematic techniques to data videos.
Second, visualization experts were interviewed about their inspirations for enhancing data storytelling from films.

\textbf{Participants.}
Our interviewees consisted of six film experts (E1-6) and five experts in data visualization experts (E7-11).
~\rv{Inspired by~\cite{krueger2014focus, lin2021engaging}, our utmost was done to diversify the backgrounds of the experts. The eleven interviewees were from the U.S. (4), Mainland China (3), Hong Kong SAR, China (3), and the U.K. (1). Seven males and four females between the ages of 25 and 48, with a minimum of two years of industry and academia experience on film or visualization, were recruited from film and visualization industries. Table \ref{tab:my-table} lists the background of the experts.

\begin{table*}
\centering
\begin{tabular}{llll}
\hline
Participant ID & Gender & Occupation(s)                                     & Experience\cr
\hline
E1             & M      & Full-time University Professor & 20 years in academia and industry\cr
\hline
E2             & M      & Senior Director                             & 10 years industry experience in film \cr
\hline
E3             & F      & Documentary Film Director                            & 9 years industry experience in documentary film  \cr
\hline
E4             & M      & Digital Visual Artist              & 5 years in both academia and industry          \cr
\hline
E5             & M      & Cinematographer                    & 5 years industry experience in cinematography \cr
\hline
E6             & M      & Visual Designer                              & 2 years industry experience in visual design \cr
\hline
E7             & M      & Researcher                              & 7 years academia experience in VIS and HCI community\cr
\hline
E8             & F      & Researcher                             & 6 years academia experience in narrative visualization \cr
\hline
E9             & M      & Researcher                              & 5 years academia experience in VIS and HCI community \cr
\hline
E10             & F      & Researcher                              & 6 years academia experience in information visualization design \cr
\hline
E11             & F      & Researcher                              & 9 years teaching experience in narrative visualization \cr
\hline
\end{tabular}
\caption[]{\rv{Background information of the 11 experts who participated in our interviews}}
\label{tab:my-table}
\vspace{-10px}
\end{table*}}


\textbf{Interview Procedure.}
~\rv{The interviews were conducted through either face-to-face or online meetings in a two-to-one manner (two authors and one expert).}
Each interview session lasted for about 120 minutes.
We first introduced our research goal, 
to help data video makers produce more understandable, impressive, and reflective endings.
We then briefed the four cinematic ending styles together with corresponding example films and data videos.
For each style,
we asked participants to give consideration and suggestions to the content and visual form of data video endings by referring to corresponding films of the same style.

\textbf{Interview Analysis}
First, three authors extracted thematic codes through an open-coding approach~\cite{charmaz2006constructing} and coded the interview transcripts separately.
Next, they iteratively coded the interview data through six rounds of meetings until reaching a consensus.
Furthermore,
an external professor in film theory and practice was requested to validate the guideline from an empirical point of view.
The whole process resulted in a final list of 20 guidelines.



\subsection{Guidelines}
\begin{table*}
\centering
\begin{tabular}{|p{0.05\textwidth} | p{0.90\textwidth}|}
\hline
                      & \multicolumn{1}{c|}{{\textbf{\textcolor{fullc}{Full Stop}}}}                                                                        \\ 
\hline
G1.1                   & Summarize facts in a declarative mood and recall the previous events of the story through montage and/or interview footage.                                            
\\ 
\hline
G1.2                   & Explicitly present the data insights or make a conclusive statement with a wide static shot.                                                        
\\ 
\hline
G1.3                   & Emphasize the connection between the data facts and the world by intercutting with real-world pictures. Engage the audiences in understanding how the real world is reflected by the data and establish emotional resonance.            \\ 
\hline
G1.4                   & Use camera effects (e.g., zoom out) to visualize the overview of data insights. 
\\ 
\hline
G1.5                   & When playing back multiple charts, juxtapose or use montage clips to arrange them.
\\ 
\hline
                      & \multicolumn{1}{c|}{\textbf{\textcolor{exclamationc}{Exclamation Point}}}                                                                                                   \\ 
\hline
G2.1                   & Present intense contradictions or surprising data facts.    
\\ 
\hline
G2.2                   & Leave audiences with a new perspective after the climax of the story.           \\ 
\hline
G2.3                   & Highlight the contradictions using colors, animations, and scenes with strong visual contrasts.     \\ 
\hline
G2.4                   & Emphasize the contradictory or surprising data facts by comparing contrasting images using visualization techniques such as juxtaposition, visual cues, and animations.                  
\\ 
\hline
G2.5                   & Use analogy to make audiences aware of the magnitude of the data to achieve surprising results.                                      \\ 
\hline
\multicolumn{1}{|c|}{} & \multicolumn{1}{c|}{\textbf{\textcolor{questionc}{Question Mark}}}                                                                                                  \\ 
\hline
G3.1                      & Animate the data change by visualizations or metaphors and end the animation abruptly to leave suspense.                \\ 
\hline
G3.2                 & Ask questions about the next movement of data, especially when the data has an incomplete action.                         
\\ 
\hline
G3.3                  & Emphasize questions (e.g., use of repeated shots) about data from new perspectives and call for actions.  
\\ 
\hline
G3.4                    & End with a perfect dream that invites or provokes audiences to question its authenticity.
\\ 
\hline
                      & \multicolumn{1}{c|}{\textbf{\textcolor{ellipsisc}{Ellipsis}}}                                                                                                         \\ 
\hline
G4.1                   & Open-end that expresses central problems unsolved with uncertainty (e.g., with gentle camera movement) and more actions are needed.                                                   
\\ 
\hline
G4.2                   & Echo the opening using similar colors, scenes, or visualizations and highlight the differences from the beginning.           
\\ 
\hline
G4.3                   & Apply the patterns of periodic data to indicate and predict possible future changes.                   \\ 
\hline
G4.4                   & Reveal data continuously using camera movement of push-in (out) and zoom-in (out) to express different possibilities for the future development.  \\ 
\hline
G4.5                   & Foreshadow the beginning of the following data video in the ending, when the topic is too large and complex to cover in a single data video.                                            
\\ 
\hline
G4.6                   & Use action or graphic match cuts to continue the ongoing topic through seamless transitions.                                  
\\ 
\hline
\end{tabular}
\caption{\label{tab:guideline}\rv{Twenty guidelines for applying the four cinematic ending styles to data videos}}
\Description{This table shows twenty guidelines for applying the four cinematic ending styles to data videos.}
\vspace{-10px}
\end{table*}

Table~\ref{tab:guideline}~provides an overview of 20 guidelines organized by each cinematic ending style (punctuation mark).


\textbf{\textcolor{fullc}{Full Stop.}}
More than half of the experts (6 out of 11) considered the full stop style useful when ``summarizing previous events at the end.''
They agreed that a full stop ending should effectively convey the take-home message. Thus, the core challenge was to make the key insight clearly understood and resonate with the audience.

They first provided suggestions on the usage scenario and content design (G1.1-1.3).
For example,
E9 commented that ``visualizations could be interpreted differently. It is important to state the take-home message explicitly and clearly at the end, which otherwise could result in a question or ellipsis mark,''
which motivated the development of G1.2.
Importantly,
experts underscored the need for developing empathy by connecting the audiences from abstract data to the real world (G1.3).
To illustrate this idea,
E2 referred to historical films,
where ``the hero/heroine often gives a monologue to summarize the past story and describe the present.'' 
E2 continued, ``the purpose is to make audiences realize that this is a true event impacting the real world, which could generate empathy and provoke thoughts.''


Experts also suggested different ways of form to deliver the content.
They (E4, E5, E7, and E8) commented that camera movements and visual cues could attract audiences' attention and emphasize important data facts,
which corresponds to G1.4.
For example, films often end with camera movements to provide a long shot, allowing audiences to take an almost ~\rv{birds-eye view.} 
They recommended different methods for reviewing multiple visualizations at the end (G1.5) because data videos often contained multiple visualizations.
Specifically,
E1 and E3 explained that montage was commonly used in films to display scenes of the past and the present and illustrate causality.
Similarly,
they advocated using montage to display multiple visualizations to illustrate their relationships.


\textbf{\textcolor{exclamationc}{Exclamation Point.}}
All experts agreed that endings with exclamation points could leave a strong impression and provided guidelines for content (G2.1-2.2) and form design (G2.3-2.5).

They first gave valuable suggestions on creating an exclamation (G2.1).
E2 and E4 commented that film endings depicted strong conflicts and contradictions to amaze the audience,
whereas E3 and E8 expressed that exclamations resulted from surprising events and data facts.
Contradictions or surprises are often intense,
creating the need for distinguishing exclamation endings from climaxes (G2.2).
``As the climax often provides the most crucial event or data findings,
the ending could show data that is sub-crucial but opens up a surprising perspective to interpret data,''
said E8.


Other experts suggested how exclamation endings could be visualized.
G2.3 was exemplified by E4's comment,
``Contradictions could be visualized following the principle of graphical design, for example, using complementary colors or shapes.''
Furthermore, they proposed using comparative visualization techniques such as animations to highlight data contrasts (G2.4).
Finally, E10 and E11 commented that audiences might not~\rv{have intuition} about the data size or magnitude.
They explained that analogy might surprise audiences by the data magnitude (G2.5).


\textbf{\textcolor{questionc}{Question Mark.}}
Most of the experts (7 out of 11) suggested that ending with a question mark was a powerful approach to provoke thoughts.
They thought highly of the film \textit{Inception} ~\cite{inception} and the data video \textit{7 Billion: How Did We Get So Big So Fast?} ~\cite{7billion} (~\autoref{fig:question}).
E6 explained, ``They both use animation to symbolize the evolution of the story and imply different endings - whether the spinning top would stop and whether the water would spout out.''
E1 further commented, ``The question arises as the storytelling is ended abruptly, but the story is still evolving.''
We derived G3.1 from those comments.


Experts described different scenarios for asking questions in data videos that informed the development of G3.2-3.4.
For example, E10 commented,
``It is common to ask questions about the future trend of the data,'' which corresponds to G3.2.
E2 and E8 considered the ``so what'' questions for calling for actions, for example, ``what should we do in response to the data facts?'' (G3.3).
Finally, E1 referred to the film endings that ``Films sometimes end with fantasy by the creators that are very different from reality, which makes audiences wonder about the reality and provoke reflections,'' which helped us develop G3.4.
 

\textbf{\textcolor{ellipsisc}{Ellipsis.}}
Seven out of 11 experts emphasized that ellipsis marks could make audiences be left wanting more,
which was the most challenging of the four styles.
Experts first advised using scenarios for ellipsis from the aspect of narration and data.

In terms of narration, experts advised two possible scenarios (G4.1-4.2).
E3 and E8 commented, ``The ending does not provide a clear answer about whether the problem at the start has been resolved, which makes audiences realize that the event is still developing.''
Their comments inspired us to develop G4.1 about content design.
G4.2 was illustrated in E2's comments, ``In films, the endings can refresh the openings by using a similar mise-en-scène to imply that the story has not ended.''
This comment echoed the feedback from data experts that ``Data videos could end with a visualization that is similar to that at the beginning with notable differences to show that data keep changing ''.
Based on their comments, we derive G4.2.




In terms of data, other experts provided suggestions based on the basis of the usage scenarios of ellipsis marks.
First,
ellipsis marks are often used to ``show the omission of repetitive or similar words'' (E5,6,9,10).
Thus,
they considered the ellipsis marks to be applicable to periodical data or data with similar patterns (G4.3).
Second,
E4 commented,
``Ellipsis marks can represent infinite, for example, using zooming in/out to change the information density.''
This idea echoes that of E11, ``such zooming brings about a feeling of  seeking answers or looking forward.''
After summarizing their comments, we derived G4.4.


Experts also mentioned series as another usage scenario of ellipsis marks in films and data videos.
E5 explained,
``In film series, the ending is often a crisis to be shown in the next film.''
Experts provided suggestions on applying this idea to data video series, especially when the dataset is large and complex. (G4.5-4.6).
Specifically,
they proposed various techniques to foreshadow the next data video, such as seamless transition techniques in films (E5) and morphing in data videos (E8).



\rv{\section{Evaluating Ending Styles and Guidelines}}
\begin{figure*}
  \includegraphics[width=\textwidth]{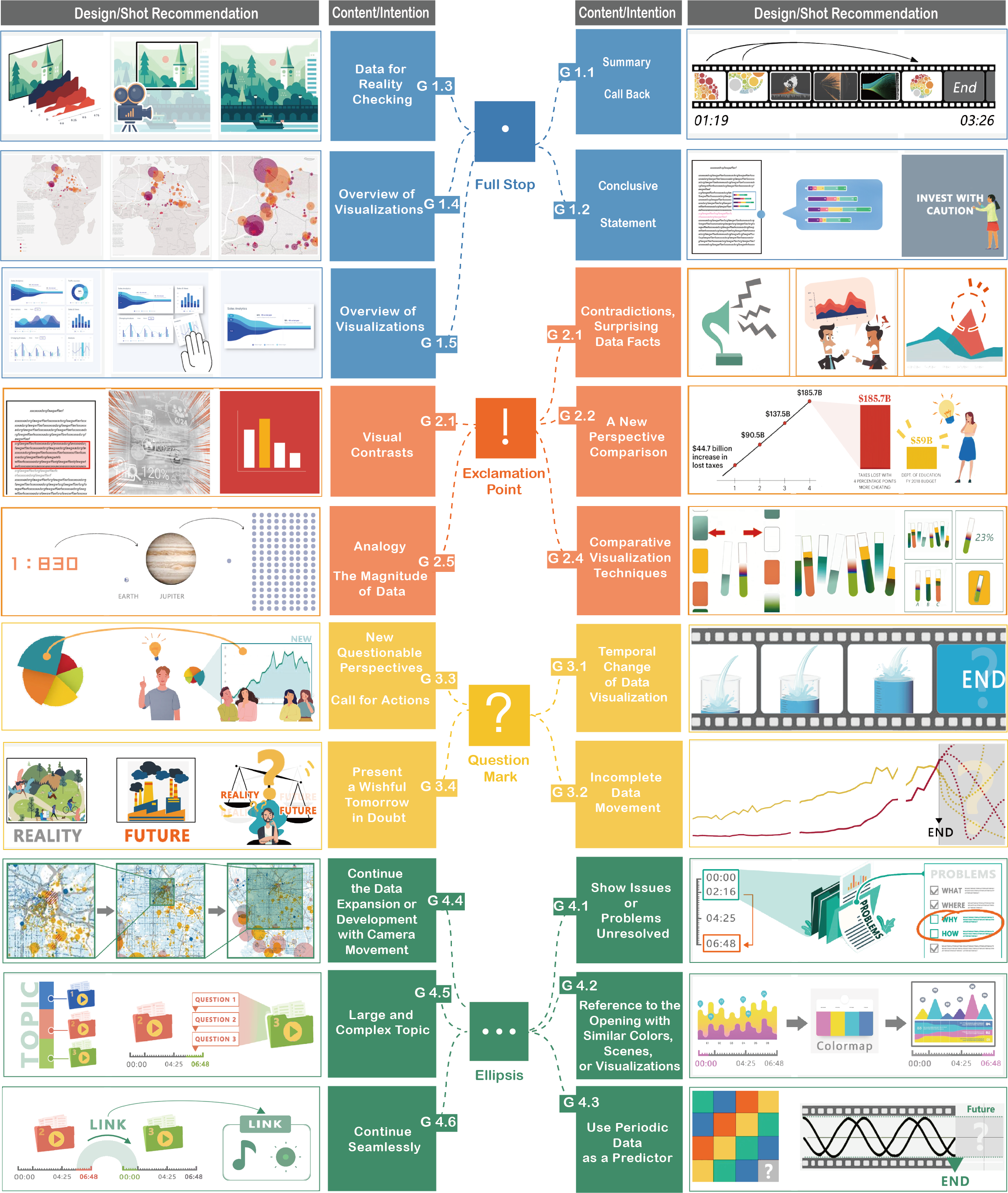}
  \caption{\rv{An overview of our guidelines includes their content/intention and design/shot recommendation. Users can select the content or intention they tend to express and then use the corresponding design or shot recommendation in the user study.}}
  \label{fig:design}
   \Description{This figure presents an overview of our twenty guidelines, including their content/intention and design/shot recommendation.
   Users can select the content or intention they tend to express and then use the corresponding design 
   or shot recommendation in the user study.}
\end{figure*}

~\rv{In this section, we conducted a user study and a comparative study to evaluate our ending styles and guidelines derived from corpus analysis and expert interviews. Specifically, we evaluated how well our ending styles and guidelines can help users design cinematic data video endings by considering two questions: (1) whether users can understand and apply the guidelines, and (2) whether the guidelines can help achieve understandable, impressive, and reflective data video endings. To answer the first question, we conducted a user study to ask \textit{guideline participants} to use and rate our endings styles and guidelines. To answer the second question, we conducted a comparative study to ask the general public and experts to rate data video endings by all participants with and without using guidelines in our user study.

Furthermore, we also presented a gallery containing four data videos that are on the same topic but end with different cinematic styles by using our guidelines in order to demonstrate the potential application of our ending styles and guidelines.} 
\subsection{User Study}

\subsubsection{Participants}
\rv{By advertising on online platforms, we recruited 24 participants (12 female) whose ages were between 24 to 31, with an average of 27.
Our participants included 16 designers, four visual effects artists, and four data analysts.
Their job involved data storytelling,
for example, delivering presentations, drafting data stories, or making data videos to analyze and communicate data.
After the user study, each participant received a remuneration of \$30 USD.}

\subsubsection{Study Materials}
We provided participants with three candidate datasets of different themes,
including \textit{the COVID-19}, \textit{the 10 causes of death}, and \textit{the global trends in over weight and obesity}.
Those datasets were selected because they were of general interest with suitable complexity.
\rv{Furthermore,
we provided participants with a sheet about our guidelines (Table~\ref{tab:guideline}), and we further explained each guideline from two aspects, their content or intention (``what'' and ``why''), and its design and shot recommendation (``how''), as shown in ~\autoref{fig:design}.
In addition, we provide participants with an interactive website (\url{https://cinematicendings.github.io/}) to browse examples of our ending styles and guidelines.}



\subsubsection{Study Design}
\rv{This user study took the between-subject protocol, that is, 12 participants were asked to create data video endings with our guidelines and 12 participants without guidelines. Specifically, we randomly and equally divided 24 participants into two groups to ensure fairness: one group with 12 participants using guidelines as~\textit{guideline participants} and the other group with 12 participants without guidelines as~\textit{non-guideline participants}. Each group had eight designers, two visual effect artists, and two data analysts. We then randomly and evenly assigned two participant groups to all four ending styles. Each ending style had three~\textit{guideline participants} and three~\textit{non-guideline participants}.
Participants could select one of the three provided datasets according to their own preferences.
The user study was hosted in a mixed mode due to the pandemic,
that is, 16 participants joined online, and eight participants were offline. They were instructed to use their familiar tools (e.g., laptops, tablets, and pens) for design. All the storyboards created by participants with or without guidelines are shown in the supplemental materials, and the 12 storyboards with our guidelines are also shown on our website (\url{https://cinematicendings.github.io/\#/Storyboard}).

\subsubsection{Study Procedure}
Our user study lasted for 55 minutes and had three phases: (1)~\textit{introduction phase}, (2)~\textit{design phase}, and (3)~\textit{feedback phase}.

\textbf{Phase 1: Introduction (around 15 minutes)}
This phase started with a brief session to introduce the study procedure, datasets, the concept of data video endings, and each cinematic ending style and guidelines to each group. For~\textit{non-guideline participants}, we introduced the same information except the guidelines.
We asked participants to create data video endings that could
foster audiences' understanding, impression, and reflection about the themes.
That said,
they were instructed to develop the core theme of their videos and subsequently design the data video endings.

\textbf{Phase 2: Design (around 25 minutes)}
Participants of both groups were given 25 minutes to craft storyboards to show the designs of their data story endings.~\textit{Guideline participants} could have our guidelines teaching materials, including guidelines (Table~\ref{tab:guideline}), design recommendation (\autoref{fig:design}), and interactive website (\url{https://cinematicendings.github.io/}).~\textit{Non-guideline participants} could search for and learn from any information from the Internet.

\textbf{Phase 3: Feedback (around 15 minutes)}
The user study ended with a 15-minute post-study interview session. All participants were asked to explain their designs. Afterward, we interviewed them to understand their considerations, challenges, and comments on creating data story endings with and without guidelines. In addition, we invited~\textit{guideline participants} to rate the understandability, impression (to what extent could the endings make a lasting impression or be memorable in the mind of the audience), reflection, clarity, and usability (to what extent could they apply the styles by using our teaching materials) of each ending style, as well as the clarity and usefulness (to what extent could the guidelines help them achieve the style) of each guideline using a 7-point Likert scale.}

\rv{\subsection{Comparative Study}
To understand whether those data story endings designed with our guidelines were truly more understandable, impressive, and reflective to audiences than those designed without guidelines, we further recruited experts from the film industry and the general public to evaluate the storyboards from our user study.
\subsubsection{Participants}
We recruited six experts and 52 general audiences. The six experts were from the film industries and different education institutions: two film directors with five and six years of experience, a cinematographer with five years of experience, a full-time university professor with 15 years of film teaching experience, one college lecturer with eight years of theater teaching experience, and one college lecturer with four years of cinematic arts teaching experience. The 52 general audiences were recruited through advertising on social media platforms and snowball sampling. They were 16 females and 36 males, and their ages were between 21 to 33, with an average age of 25.}

\subsubsection{Study Design and Procedure}
We invited 52 general audiences to assess storyboards according to three criteria that conformed to our research goals,
i.e., whether the storyboard can enhance the understandability, impression, and reflection of the core theme.
Six experts were asked to provide a professional assessment on two \rv{more} criteria that required expertise, 
including creativeness \rv{(i.e., to what extent these storyboards be indicated as original and unique)} and \rv{cinematic expression (i.e., to what extent these storyboards could be regarded as cinematic).}
The experts and general audiences rated all 24 storyboards on a 7-point Likert scale. 
Storyboards were grouped according to the same ending style to ensure comparability,
whereas the order was randomly shuffled for each judge to reduce biases.
Judges were blind to the control conditions (i.e., the existence of design guidelines).

\begin{figure}[!t]
\centering
\includegraphics[width=\linewidth]{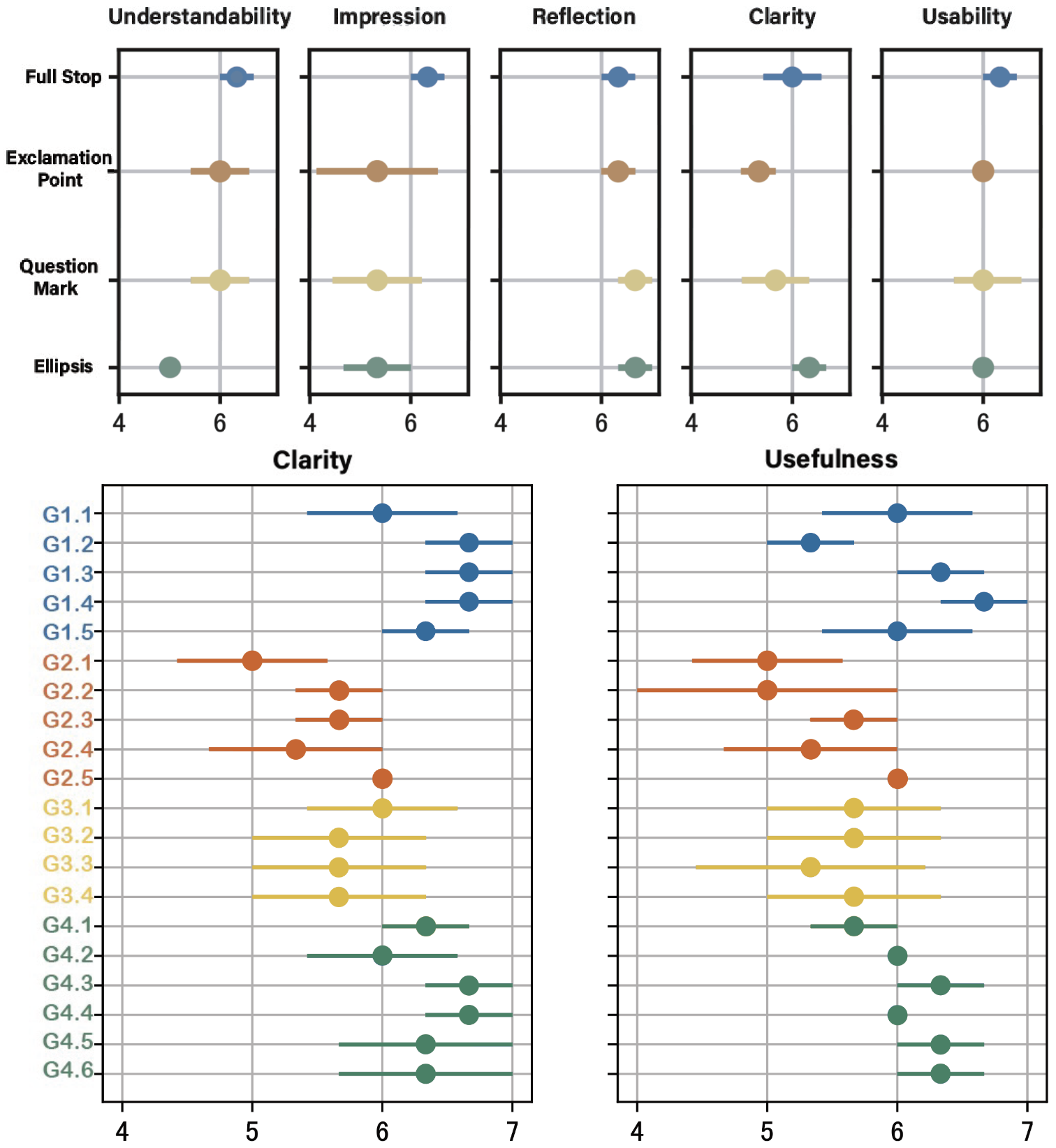}
  \caption{\rv{Guideline participants' ratings with standard errors on the punctuation marks and guidelines. The upper part shows their ratings on each punctuation mark. The lower part shows their rations on each guideline.}}
  \Description{This figure shows the guideline participants' ratings on the punctuation marks and guidelines. For each punctuation, the specific average scores include: (1) the understandability of Full Stop ($Mean = 6.33, SD = 0.58$), Exclamation Point ($Mean = 6.00, SD = 1.00$), Question Mark ($Mean = 6.00, SD = 1.00$), Ellipsis ($Mean = 5.00, SD = 0.00$), (2) the impression of Full Stop ($Mean = 6.33, SD = 0.58$), Exclamation Point ($Mean = 5.33, SD = 2.08$), Question Mark ($Mean = 6.33, SD = 1.53$), Ellipsis ($Mean = 5.33, SD = 1.15$), (3) the reflection of Full Stop ($Mean = 6.33, SD = 0.58$), Exclamation Point ($Mean = 6.33, SD = 0.58$), Question Mark ($Mean = 5.67, SD = 0.58$), Ellipsis ($Mean = 6.67, SD = 0.58$), (4) the clarity of Full Stop ($Mean = 6.00, SD = 1.00$), Exclamation Point ($Mean = 5.33, SD = 0.58$), Question Mark ($Mean = 5.67, SD = 1.15$), Ellipsis ($Mean = 6.33, SD = 0.58$), and (5) the usability of Full Stop ($Mean = 6.00, SD = 0.58$), Exclamation Point ($Mean = 6.00, SD = 0.00$), Question Mark ($Mean = 6.33, SD = 1.00$), Ellipsis ($Mean = 6.00, SD = 0.00$).
For each guideline, the specific average scores include: (1) the clarity of G1.1 ($Mean = 6.00, SD = 1.00$), G1.2 ($Mean = 6.67, SD = 0.58$), G1.3 ($Mean = 6.67, SD = 0.58$), G1.4 ($Mean = 6.67, SD = 0.58$), G1.5 ($Mean = 6.33, SD = 0.58$), G2.1 ($Mean = 5.00, SD = 1.00$), G2.2 ($Mean = 5.67, SD = 0.58$), G2.3 ($Mean = 5.67, SD = 0.58$), G2.4 ($Mean = 5.33, SD = 1.15$), G2.5 ($Mean = 6.00, SD = 0.00$), G3.1 ($Mean = 6.00, SD = 1.00$), G3.2 ($Mean = 5.67, SD = 1.15$), G3.3 ($Mean = 5.67, SD = 1.15$), G3.4 ($Mean = 5.67, SD = 1.15$), G4.1 ($Mean = 6.33, SD = 0.58$), G4.2 ($Mean = 6.00, SD = 1.00$), G4.3 ($Mean = 6.67, SD = 0.58$), G4.4 ($Mean = 6.67, SD = 0.58$), G4.5 ($Mean = 6.33, SD = 0.58$), G4.6 ($Mean = 6.33, SD = 1.15$), and (2) the usefulness of G1.1 ($Mean = 6.00, SD = 1.00$), G1.2 ($Mean = 5.33, SD = 0.58$), G1.3 ($Mean = 6.33, SD = 0.58$), G1.4 ($Mean = 6.67, SD = 0.58$), G1.5 ($Mean = 6.00, SD = 1.00$), G2.1 ($Mean = 5.00, SD = 1.00$), G2.2 ($Mean = 5.00, SD = 1.73$), G2.3 ($Mean = 5.67, SD = 0.58$), G2.4 ($Mean = 5.33, SD = 1.15$), G2.5 ($Mean = 6.00, SD = 0.00$), G3.1 ($Mean = 5.67, SD = 1.15$), G3.2 ($Mean = 5.67, SD = 1.15$), G3.3 ($Mean = 5.33, SD = 1.53$), G3.4 ($Mean = 5.67, SD = 1.15$), G4.1 ($Mean = 5.67, SD = 0.58$), G4.2 ($Mean = 6.00, SD = 0.00$), G4.3 ($Mean = 6.33, SD = 0.58$), G4.4 ($Mean = 6.00, SD = 0.00$), G4.5 ($Mean = 6.33, SD = 0.58$), G4.6 ($Mean = 6.33, SD = 0.58$).}
  \label{fig:workshop_result}
\end{figure}

\begin{figure*}[!t]
\centering
\includegraphics[width=\textwidth]{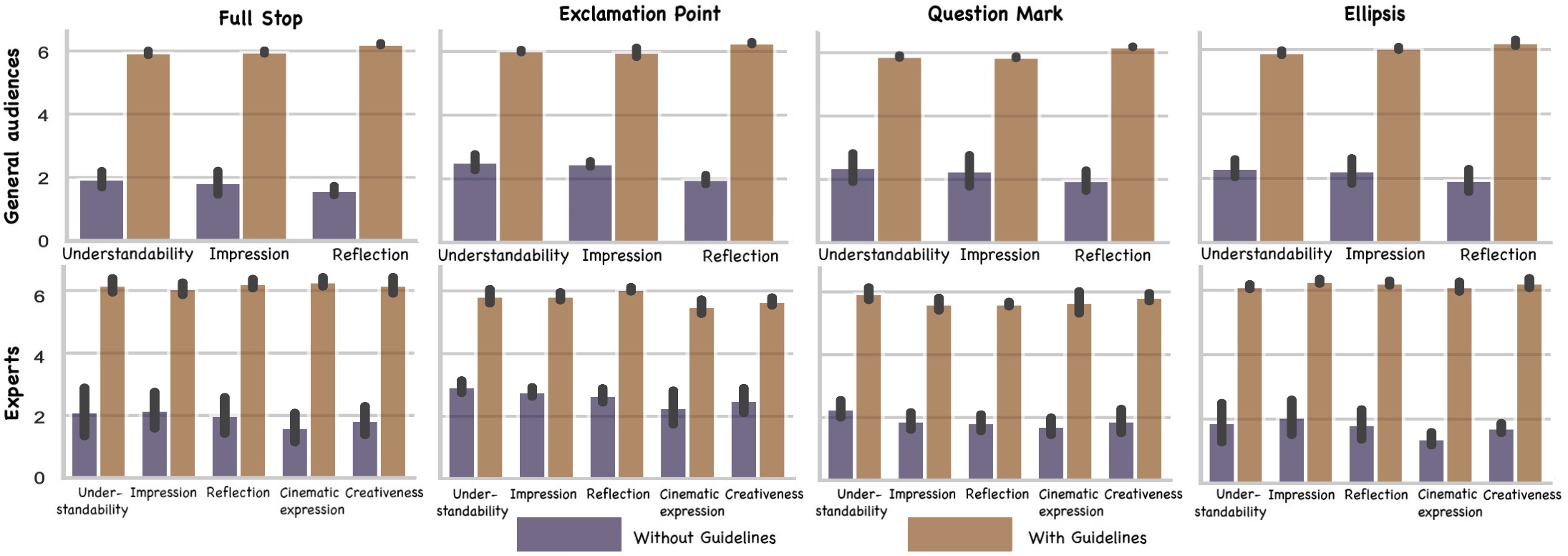}
  \caption{\rv{Experts and the general audiences' ratings on the storyboards with standard errors. The upper chart shows the scores of understandability, impression, and reflection about the storyboard ratings by the general audiences, and the lower chart shows the scores of understandability, impression, reflection, cinematic expression, and creativeness about the storyboard ratings by experts.}}
  \Description{This figure shows the experts' and the general audiences' ratings on storyboards with and without guidelines. The upper chart shows the scores of understandability, impression, and reflection about the storyboard ratings by the general audiences, including 
  (1)with guidelines: understandability for full stop ($Mean = 5.94, SD = 0.10$), impression for full stop ($Mean = 5.97, SD = 0.08$), reflection for full stop ($Mean = 6.22, SD = 0.06$); understandability for exclamation ($Mean = 6.01, SD = 0.06$), impression for exclamation ($Mean = 5.97, SD = 0.19$), reflection for exclamation ($Mean = 6.26, SD = 0.06$); understandability for question ($Mean = 5.88, SD = 0.06$), impression for question ($Mean = 5.85, SD = 0.05$), reflection for question($Mean = 6.18, SD = 0.03$); understandability for ellipsis ($Mean = 5.90, SD = 0.10$), impression for ellipsis ($Mean = 6.04, SD = 0.07$), reflection for ellipsis ($Mean = 6.21, SD = 0.13$); 
  (2)without guidelines: understandability for full stop ($Mean = 1.94, SD = 0.34$), impression for full stop ($Mean = 1.83, SD = 0.48$), reflection for full stop ($Mean = 1.58, SD = 0.20$); understandability for exclamation ($Mean = 2.52, SD = 0.32$), impression for exclamation ($Mean = 2.47, SD = 0.12$), reflection for exclamation ($Mean = 1.97, SD = 0.19$); understandability for question ($Mean = 2.37, SD = 0.57$), impression for question ($Mean = 2.27, SD = 0.62$), reflection for question($Mean = 1.96, SD = 0.41$); understandability for ellipsis ($Mean = 2.31, SD = 0.36$), impression for ellipsis ($Mean = 2.23, SD = 0.50$), reflection for ellipsis ($Mean = 1.94, SD = 0.45$).
  The lower chart shows the scores of understandability, impression, reflection, cinematic expression, and creativeness about the storyboard ratings by experts, including 
  (1)with guidelines: understandability for full stop ($Mean = 6.17, SD = 0.29$), impression for full stop ($Mean = 6.06, SD = 0.25$), reflection for full stop ($Mean = 6.22, SD = 0.19$), cinematic expression for full stop ($Mean = 6.28, SD = 0.19$), creativeness for full stop ($Mean = 6.17, SD = 0.33$); understandability for exclamation ($Mean = 5.83, SD = 0.29$), impression for exclamation ($Mean = 5.83, SD = 0.17$), reflection for exclamation ($Mean = 6.06, SD = 0.10$), cinematic expression for exclamation ($Mean = 5.50, SD = 0.29$), creativeness for exclamation ($Mean = 5.67, SD = 0.44$); understandability for question ($Mean = 5.94, SD = 0.25$), impression for question ($Mean = 5.61, SD = 0.25$), reflection for question($Mean = 5.61, SD = 0.10$), cinematic expression for question ($Mean = 5.67, SD = 0.44$), creativeness for question ($Mean = 5.83, SD = 0.17$); understandability for ellipsis ($Mean = 6.11, SD = 0.10$), impression for ellipsis ($Mean = 6.28, SD = 0.10$), reflection for ellipsis ($Mean = 6.22, SD = 0.10$), cinematic expression for ellipsis ($Mean = 6.11, SD = 0.19$), creativeness for ellipsis ($Mean = 6.22, SD = 0.19$); 
  (2)without guidelines: understandability for full stop ($Mean = 2.11, SD = 0.96$), impression for full stop ($Mean = 2.17, SD = 0.73$), reflection for full stop ($Mean = 2.00, SD = 0.73$), cinematic expression for full stop ($Mean = 1.61, SD = 0.59$), creativeness for full stop ($Mean = 1.83, SD = 0.58$); understandability for exclamation ($Mean = 2.94, SD = 0.25$), impression for exclamation ($Mean = 2.77, SD = 0.19$), reflection for exclamation ($Mean = 2.67, SD = 0.29$), cinematic expression for exclamation ($Mean = 2.28, SD = 0.67$), creativeness for exclamation ($Mean = 2.50, SD = 0.50$); understandability for question ($Mean = 2.28, SD = 0.35$), impression for question ($Mean = 1.89, SD = 0.35$), reflection for question($Mean = 1.83, SD = 0.33$), cinematic expression for question ($Mean = 1.72, SD = 0.35$), creativeness for question ($Mean = 1.89, SD = 0.48$); understandability for ellipsis ($Mean = 1.89, SD = 0.75$), impression for ellipsis ($Mean = 2.06, SD = 0.67$), reflection for ellipsis ($Mean = 1.83, SD = 0.58$), cinematic expression for ellipsis ($Mean = 1.39, SD = 0.25$), creativeness for ellipsis ($Mean = 1.39, SD = 0.25$).}
  \label{fig:comparative_result}
  \vspace{-10px}
\end{figure*}

\rv{\subsection{User Study and Comparative Study Results}}
\rv{In this section, the quantitative and qualitative feedback from the participants in the user study and comparative study are reported.}
\rv{\subsubsection{User Study Results}

Generally, participants who used guidelines agreed on their satisfaction ($M = 5.5, SD = 1.31$) with their designs as well as the inspirations ($M = 6, SD = 1.04$) our guidelines provided on a 7-point Likert scale.}

With our teaching materials, they found that they could clearly understand \rv{($M = 5.83, SD = 0.83$)} and use \rv{($M = 6.08, SD = 0.51$)} each punctuation with corresponding guidelines. 
They also provided positive feedback on the overall effects of each punctuation with corresponding guidelines in enhancing the understandability \rv{($M = 5.83, SD = 0.83$)}, impression \rv{($M = 5.58, SD = 1.31$)}, and reflection \rv{($M = 6.5, SD = 0.52$)} of the core theme.
For clarity and usefulness (to what extent
the guideline could help participants realize punctuation), the participants
provided positive scores (\autoref{fig:workshop_result}).

\rv{In the post-study interview, the main consideration for endings' design was how to express data insight completely and intuitively, which could be facilitated by our cinematic styles and guidelines. P8 commented, ``The idea of four punctuation marks as a classification and framework of different ending contextualize both visual and content, which is very intuitive and impressive.''
All participants with guidelines (P1-P12) agreed that our guidelines were diverse and helpful to inspire their creating data video endings. P12 praised the inspiration of G4.4 ``the shot design recommendation of zoom-in (out) to present the data expansion or development inspires my ellipsis ending design a lot while I was thinking about how to extend the new information more fluently.'' 
However, they suggested that our guidelines could add considerations about audio in the future.}


\rv{\subsubsection{Comparative Study Results}
For each storyboard, we averaged the score of the five experts on how much it enhanced the understandability, impression, and reflection of the theme as well as its cinematic expression and creativeness (\autoref{fig:comparative_result}).
For each metric, we conducted a t-test on the score of storyboards designed with and without guidelines.
We found that the scores of storyboards using the guidelines were significantly higher than those of storyboards not using guidelines (all $ p < 0.001 $) on all five metrics.
From the rating by the general public, we found similar results.
In terms of the metrics---understandability, impression, and reflection, the scores of storyboards guided by guidelines were significantly higher than those of storyboards without guidelines (all $ p < 0.001 $).}

\subsection{Gallery}
\begin{figure*}[!t]
\centering
\includegraphics[width=\textwidth]{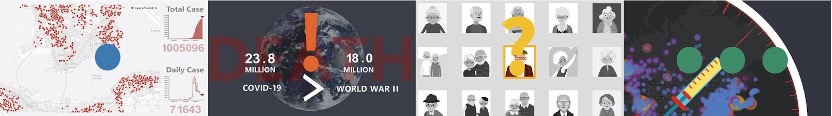}
  \caption{\rv{Four data video endings that applied our four cinematic ending styles and guidelines. From left to right, four screenshots of four data video endings with \textit{full stop}, \textit{exclamation point}, \textit{question mark}, and \textit{ellipsis} are from the gallery on our website.}}
  \Description{This figure shows four data video endings that used our four cinematic ending styles and guidelines. From left to right, four screenshots of four data video endings with \textit{full stop}, \textit{exclamation point}, \textit{question mark}, and \textit{ellipsis} are from the gallery on our website.}
  \label{fig:fourendings}
  \vspace{-10px}
\end{figure*}

To demonstrate the application of our cinematic guidelines, three authors of this work (a film scriptwriter, a film visual designer, and a data visualization designer) created four data videos\rv{, each of which ended on one of} the four punctuation marks (\autoref{fig:fourendings}). 
The data videos were about the recent COVID-19 pandemic and demonstrate how each of the four ending punctuation marks can be visualized with different cinematic ending styles. 
For example, based on the ellipsis ending style and corresponding guidelines G4.1, G4.2, and G4.4, we have designed a visualization of a running time clock and the fading images of the elderly, symbolizing and emphasizing the great sense of urgency to improve the vaccination rate. 
These videos can be found in the supplementary materials and on our website (\url{https://cinematicendings.github.io/}).

\section{Discussion}

\rv{In this section,
we discuss the limitation of our work and the lessons learned for improving data storytelling.}

\textbf{Towards a deeper understanding of data story endings.}
We classified data story endings according to the framework of punctuation marks,
which are intuitive enough to understand and help communicate the relevance between films and data videos.
However, our classification was not without limitations.
First, punctuation marks can be used in combination.
For instance, the interrobang mark\footnote{\url{https://en.wikipedia.org/wiki/Interrobang}} is a common technique in advertisement and comic endings to combine the functions of the question and the exclamation mark.
In the future, we plan to study such combined use to deepen our understanding of data story endings.
Second, different punctuation marks might generate similar effects on audiences despite their differences in form and content design.
For instance, both ellipsis and question mark could leave audiences in suspense, whereas the former has a smaller, softer effect.
To distinguish them better, we plan to investigate their influences on audiences.
~\rv{In addition, data analytics and data storytelling should be balanced and incorporated in the data story ending. 
Data and analytic thinking might be decoupled from thoughts and feelings. However, it is crucial to follow the data analysis process to conclude data insights~\cite{provost2013data} for data story endings in the first step and then utilize storytelling techniques to help reveal information and data insights effectively and intuitively~\cite{gershon2001storytelling}. 
The importance of balancing data analytics and storytelling is recognized, and we plan to observe and explore their relationship further in future work.}

\textbf{Making guidelines visual and accessible.}
A core challenge encountered during this research was to make guidelines more accessible,
that is, simple and intuitive enough for designers to understand and appreciate easily.
Our solutions were the ``visualization'' of guidelines that helps designers browse and select guidelines based on their intended styles (\autoref{fig:design}).
In addition to text descriptions,
we created intuitive diagrams visualizing each guideline.

Furthermore,
an interactive website (\url{https://cinematicendings.github.io/}) was provided for designers to explore examples,
facilitating their comprehension and understanding.
The participants in our experiment highly appreciated those materials.
To benefit future researchers and users,
we created data videos of four different ending styles based on our guidelines.
We also placed those videos together with participants' storyboards in our online gallery to promote research in narrative visualizations by combining visual and storytelling techniques in cinematic arts, and propose practical guidelines for designers to improve their data storytelling ``in the wild.''
Moving forward,
we plan to continue research on integrating film and visualization studies and investigating other factors such as emotions.

\rv{\textbf{Creating interactive tools for data storytelling.}
During our user study, participants were asked about their expectations of future tools for creating data stories. Participants praised the usefulness of our interactive website and study materials, which guided them in selecting punctuation marks, shot/design recommendations, and guidelines step by step. Specifically, they hoped for an interactive system that could recommend narration structures given the user-input data tables and constraints (e.g., ``ending with a question mark''). Such intelligent approaches could help facilitate the design process in crafting story structures and improve efficiency and accuracy. 
Future work can further combine these intelligent approaches with clear story structure temples as a fundamental part of interactive data video authoring tools.}


\rv{\textbf{Generalizability of the Cinematic Styles and Guidelines.}
We studied the endings of narrative visualizations in the form of data videos. Our cinematic styles and guidelines could be extended to other narrative visualization forms~\cite{segel2010narrative, ragan2015characterizing}. 
First, our cinematic styles and guidelines can be applicable to non-interactive narrative visualization that consists of a sequence of frames similar to the key property of data videos to present their data insights, such as \textit{data GIFs}. 
For example, our ending style ``exclamation point'' and its guidelines are suitable for \textit{data GIFs} which are expected to convey messages in a short time~\cite{shu2020makes} and leave the audience with an impression and surprise.
Second, our cinematic styles and guidelines can be generalized to interactive visualization endings. 
Inspired by interactive films~\cite{clarke2001film} and interactive data comics~\cite{wang2021interactive}, data videos can be ended with multiple endings to create different effects. 
In our gallery, we have created four data videos that convey the same data insight but end with four types of cinematic endings (\autoref{fig:fourendings}).
Thus, we propose that four types of endings can be chosen by designers for creating interactive data stories which could present different endings based on audiences’ selections.
This interactive structure is similar to the Martini Glass visualization structure~\cite{segel2010narrative}. The audience should follow the author-driven narrative in the opening and climax~\cite{freytag1908freytag} of data videos, and then they can interact with the multiple endings through the reader-driven narrative. Once the audience selects one of the ending styles based on their preferences, this ending may potentially become a jumping off point for the next data visualization with new insights. Therefore, our diverse data video ending designs show a potential form of interactive visualization applications. 
In addition, these ending styles remain unclear on how to interact appropriately with audiences’ preferences. We can systematically investigate the interactive ending designs based on audiences’ preferences in future work.}

\section{Conclusion}
Endings have received wide attention in literature and film studies.
Endings are not the real end of the story but instead emphasize the concluding statement with stronger impressions and persuasion powers.
We studied how to enhance the endings of data videos by drawing inspiration from hundreds of~\rv{famous} film endings and contextualizing them to popular data videos.
Our work highlights an interdisciplinary methodology,
that is,
we involved participants from different backgrounds (e.g.,~film experts, designers, and data analysts) in all stages of our work (e.g., \rv{analyzing films and data videos to derive styles, deriving guidelines through expert interviews, and validating guidelines through a user study and a comparative experiment}).
We hope our work could inspire future research to enhance data storytelling by extending not only the disciplinary depth but also the interdisciplinary breadth. 


\begin{acks}
This research was supported in part by Hong Kong Theme-based Research Scheme T41-709/17N. We would like to thank our interview and user study participants for sharing their thoughts with us. Finally, we thank our reviewers for their constructive comments.

\end{acks}

\balance





\providecommand{\noopsort}[1]{}




\end{document}